\documentclass[usegraphicx,useAMS,usenatbib]{mn2e}

\newcommand{\kms}{\,km\,s$^{-1}$}

\newcommand{\be}{\begin{equation}}
\newcommand{\ee}{\end{equation}}
\newcommand{\bd}{\begin{displaymath}}
\newcommand{\ed}{\end{displaymath}}

\title[The distance to the ONC]
  {The distance to the Orion Nebula
  Cluster}
\author[R.D. Jeffries]
  {R.D.~Jeffries\\
  Astrophysics Group, School of Physical and Geographical Sciences, Keele University, Keele, 
      Staffordshire ST5 5BG\\
}
\date{Submitted December 5 2006}

\pagerange{\pageref{firstpage}--\pageref{lastpage}} \pubyear{2004}

\def\LaTeX{L\kern-.36em\raise.3ex\hbox{a}\kern-.15em
    T\kern-.1667em\lower.7ex\hbox{E}\kern-.125emX}

\begin{document}

\label{firstpage}

\maketitle

\begin{abstract}
The distance to the Orion Nebula Cluster (ONC) is estimated using the
rotational properties of its low-mass pre main-sequence (PMS)
stars. Rotation periods, projected equatorial velocities and
distance-dependent radius estimates are used to form an observational
$\sin i$ distribution (where $i$ is the axial inclination), which is
modelled to obtain the distance estimate.  A distance of $440\pm34$,pc
is found from a sample of 74 PMS stars with spectral types between G6
and M2, but this falls to $392\pm 32$\,pc when PMS stars with accretion
discs are excluded on the basis of their near-infrared excess. Since
the radii of accreting stars are more uncertain and probably
systematically underestimated, then this closer distance is preferred.
The quoted uncertainties include statistical errors and uncertainties
due to a number of systematic effects including binarity and
inclination bias.  This method is geometric and independent of stellar
evolution models, though does rely on the assumption of random axial
orientations and the Cohen \& Kuhi (1979) effective temperature scale
for PMS stars. The new distance is consistent with, although lower and
more precise, than most previous ONC distance estimates. A closer ONC
distance implies smaller luminosities and an increased age based on the
positions of PMS stars in the Hertzsprung-Russell diagram.
\end{abstract}

\begin{keywords}
 stars: formation -- stars: distances -- methods: statistical -- open
 clusters and associations: M42 
\end{keywords}

\section{Introduction}

The Orion Nebula cluster (ONC) is among the best studied star
forming regions. It is perhaps the premier cluster for investigating
star formation and early stellar evolution because it is relatively
nearby, very young ($<2$\,Myr) and contains a large population
of stars and brown dwarfs covering the entire (sub)stellar
mass spectrum ($0.01<M/M_{\odot}<30$ -- see Hillenbrand 1997; Slesnick,
Hillenbrand \& Carpenter 2004).  The ONC lies just
in front of the dense OMC-1 molecular cloud, so background contamination
of cluster candidates is small and proper motion studies have
successfully ascribed membership to almost 1000 stars (Jones \&
Walker 1988). The ONC is a focus for understanding
the initial stellar mass function, the evolution of circumstellar
discs, early stellar angular momentum loss, the influence of
high mass stars on lower mass siblings and their discs, X-ray activity
in young stars, the history of star formation and the formation of star
clusters in general (e.g. see Hillenbrand et al. 1998; O'Dell 1998;
Lada et al. 2000; Muench et al. 2002; Herbst et al. 2002; Preibisch et
al. 2005; Huff \& Stahler 2006; Shuping et al. 2006 among many others).

The distance to
the ONC is quite poorly constrained -- anywhere between 350\,pc and
550\,pc, with a most likely range of 400--500\,pc depending on which
techniques are considered most reliable (see section~2). 
The distance is an important
parameter in determining absolute dimensions, velocities and mass loss
rates in the cluster as well as for estimating stellar luminosities and
hence masses and ages from evolutionary tracks.  

In this paper the distance to the ONC is determined using the
rotational properties of pre main-sequence (PMS) stars (axial rotation
periods $P$, and projected equatorial velocities $v\sin i$, where $i$
is the inclination angle of the spin axis to the line of sight),
together with estimates of stellar radii. From these, the observed
distribution of $\sin i$ can be compared to a model that assumes random
spin axis orientation and appropriate contributions from uncertainties
in the observational parameters. Because the estimated stellar radii
scale linearly with distance, the distance can be treated as a free
parameter in the model to optimise the match between observed and
predicted distributions.

This technique was developed initially by Hendry,
O'Dell \& Collier Cameron (1993) and used
to derive a distance of $132\pm 10$\,pc to the Pleiades, which
compares well with the currently accepted Pleiades distance from
numerous other techniques (O'Dell, Hendry \& Collier Cameron 1994).
Preibisch \& Smith (1997) used a variant of the method to
find a distance to T-Tauri stars in the Taurus star forming
region. Again, the distance they found of $152\pm 10$\,pc now compares
well with distances from Hipparcos parallaxes (Bertout, Robichon \&
Arenou 1999).

In section 2 previous estimates of the distance to the ONC are
reviewed. The database of ONC rotation measurements that were used for
this work is discussed in section 3.  The modelling technique is
described in section 4, which also discusses systematic effects that
must be taken into account. The results are presented in section 5 and
their robustness against variations in the model parameters and
assumptions is tested. Conclusions and a brief discussion are presented
in section 6.

\section{Previous distance estimates to the ONC}

Most previous distance estimates were based on fits to the upper main
sequence of the ONC in the Hertzsprung-Russell (HR) diagram and
hampered by colour excesses, variable extinction, the nebular
background and uncertain binarity. Walker (1969) found a distance
modulus of $8.37\pm 0.05$ (472\,pc) from $UBV$ photometry of relatively
unobscured stars in the outskirts of the ONC. However, Penston, Hunter
\& O'Neill (1975) obtained optical and near infrared photometry in a 40
arcminute box around the ONC, determining distance moduli of
$8.1\pm0.1$ (417\,pc) from the $V$,$B-V$ colour magnitude diagram
(CMD), and only 7.7 (347\,pc) from the $V$,$V-I$ CMD. More recent
estimates used the large $UBVubvy\beta$ dataset of Warren \& Hesser
(1978). These authors found a distance modulus of $8.42\pm0.24$
(483\,pc) for the Orion OB1d1 subregion, which is a $\sim 30$ arcminute
diameter box around the ONC, but excluding the central few arcminutes
around the central Trapezium cluster of high-mass stars. Anthony-Twarog
(1982) used the same data with somewhat different calibrations to
obtain a distance modulus of $8.19\pm 0.10$ (435\,pc). Breger, Gehrz \&
Hackwell (1981) used infrared and broad band polarisation measurements
to exclude objects that showed abnormal reddening or evidence for
circumstellar material and found that the remaining objects have an
average distance modulus of $8.0\pm 0.1$ (400\,pc) in the $V$,$B-V$
CMD.  The significant discrepancy between for example Walker's (1967)
and Breger et al.'s (1981) distance estimates points to the fact that
there may be substantial additional systematic uncertainties present in
these HR diagram-based results.

Unfortunately, no light is shed on these varying distance
estimates by other independent techniques. 
Only one star in the ONC has a direct parallax
measurement from Hipparcos, 
yielding a distance of $361^{+168}_{-87}$\,pc (Bertout et
al. 1999). A more indirect but frequently cited constraint was
provided by Genzel et al. (1981). Using an expanding cluster
parallax method they analysed the proper motions and radial
velocities of H$_2$O masers in the Kleinmann-Low nebula (part of
OMC-1), thought to be just behind the ONC, and
obtained a distance of $480\pm 80$\,pc.  Finally, Stassun et al. (2004)
found an eclipsing binary just 20 arcminutes south
of the Trapezium stars. They determined absolute radii and effective
temperatures for the components and hence estimated a
distance of $419\pm 21$\,pc from the system photometry. However, Stassun et
al. suggest that this binary system might be foreground to the
ONC because at that distance the object appears older by a few Myr
than the bulk of the ONC population.

\section{Construction of the database}

\begin{figure}
\includegraphics[width=75mm]{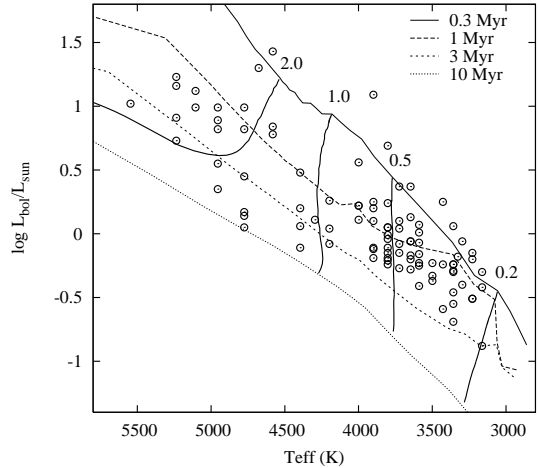}
\caption{A Hertzsprung-Russell diagram for the 95 stars of the ONC
  rotation sample (see text) assuming an ONC distance of 470\,pc.
Model isochrones and evolutionary tracks from Siess et al. (2000) are
  shown. Masses are labelled in solar units.
}
\label{hrdiag}
\end{figure}

The observed $\sin i$ distribution in the ONC was investigated using a
database founded on the catalogue of ONC axial rotation periods
constructed by Herbst et al. (2002). This catalogue also includes
rotation periods measured by Stassun et al. (1999) and Herbst et
al. (2000).  From this, all objects were selected that had an entry in
the Jones \& Walker (1988) catalogue of proper motions within 15
arcminutes of the ONC centre and a spectral type, effective temperature
($T_{\rm eff}$), luminosity ($L$) and radius ($R$) estimate listed by
Hillenbrand (1997). The $L$ and $R$ estimates in Hillenbrand (1997)
assume an ONC distance of 470\,pc.  Almost all of these stars also have
a measurement of their $I-K$ excess -- i.e. the excess colour over that
expected from the photosphere of a ``normal'' star of similar spectral
type.  Measurements of the equivalent width (EW) of the Ca\,{\sc ii}
8542\AA\ line, which is also diagnostic of accretion, are also
available for most of the sample from Hillenbrand et al. (1998).

From this subset, $v \sin i$ measurements were found in the catalogues
of Rhode et al. (2001) and Sicilia-Aguilar et al. (2005). The latter
work was performed at a higher resolving power (34\,000 versus 21\,500)
and so values from Sicilia-Aguilar et al. were preferred for $v \sin i$
values less than 20\kms\ .  For higher velocities the measurement with
the best quoted fractional precision was chosen.  Objects with only
upper limits to their $v\sin i$ were excluded.  Three objects (JW 192,
JW 381 and JW 710) where Herbst et al. (2002) state that the measured
periods are highly uncertain were also excluded.

\begin{table*}
\caption{Database of objects considered in this work. The full table
  is available in electronic form and contains 95 rows. A sample is
  given here to illustrate its content. Column 1 gives
the identifying number from Jones \& Walker (1988); columns 2 and 3 list the
spectral type and $T_{\rm eff}$ from Hillenbrand (1997); columns 4, 5
and 6 list the $v \sin i$, its uncertainty and the source of the
rotational velocity membership (1 -- Rhode et al. [2001], 2 --
Sicilia-Aguilar et al. [2005]); column 7 lists the rotational period as
given by Herbst et al. (2002); columns 8, 9 and 10 list the $\log$
bolometric luminosity (in solar units), stellar radius (assuming a
cluster distance of 470\,pc)
and $\Delta (I-K)$ from Hillenbrand (1997); column 11 gives the
  equivalent width of the 8542\AA\ Ca\,{\sc II} line from Hillenbrand
  et al. (1998) [negative indicates an emission line]; column 12 lists the derived
$\sin i$.} 
\begin{tabular}{cccccccccccc}
\hline
  JW  &    SpT & $T_{\rm eff}$ & $v\sin i$ & $\Delta v\sin i$ & Source & $P$ &
  $\log L/L_{\odot}$ & $R/R_{\odot}$ & $\Delta (I-K)$ & EW[Ca] & $\sin i$ \\
      &        &  (K)     & (\kms)  & (\kms) & & (d) & & & (mag) & (\AA) &\\
\hline
   3  &     K8 & 3801.8 &  29.8 &  4.5 & 1 &  3.43 &   0.24 & 3.036 &   0.55 &2.0& 0.665  \\
  17  &     M4 & 3228.4 &  19.9 &  3.9 & 1 &  3.14 &  -0.20 & 2.537 &   0.10 &0.8& 0.486  \\
  20  &   M3.5 & 3296.0 &  70.3 & 11.6 & 1 &  0.67 &  -0.40 & 1.933 &   0.10 &0.7& 0.481  \\
  25  &   M4.5 & 3162.2 &  18.0 &  4.0 & 1 &  2.28 &  -0.88 & 1.209 &   0.52 &0.0& 0.671  \\
\hline
\end{tabular}
\label{database}
\end{table*}

\subsection{Measurement uncertainties}

The periods measured for low-mass PMS stars are generally very precise
and accurate, relying on the co-rotation of magnetic inhomoheneities
(starspots) in the photosphere. Providing a sufficient time baseline
and sampling frequency are obtained, period precisions of better than 1
per cent are usually achieved. Studies of objects with periods measured
at more than one epoch suggest that $\simeq
90$ per cent of period measurements have at least this level of precision
(Herbst et al. 2002).  About 10 per cent of period measurements can be
catastrophically incorrect. The periods could be too short by a factor
of two if more than one spotted region on a star leads to a
``double-humped'' light curve that is interpreted
incorrectly. Alternatively, the measured period can be {\em much}
longer than the true periods because of aliasing with the $\simeq$1 day
sampling interval that is present in almost all datasets (see
discussion in Herbst et al. 2002). Many of these incorrect periods have
already been weeded out by Herbst et al. (2002) and others are
suspected (e.g. the three stars excluded in the previous subsection),
but it is possible that a small fraction remain. For now an
uncertainty of 1 per cent on all the period measurements is assumed.

The resolution of the rotational velocity studies limits the $v \sin i$
measurements to values of more than 11\kms\ and about 5\kms\ from Rhode
et al. (2001) and Sicilia-Aguilar et al. (2005) respectively.  Above
these limits the fractional precision of the $v \sin i$ measurements
are of order 10 per cent, though vary from object-to-object. Rhode et
al. (2001) discuss how problems in calibrating the intrinsic resolution
of fibre spectrographs can lead to problematic and possibly incorrect
values for $v \sin i$ when rotational broadening is close to the
detection threshold. Rhode et al. suggest using $v \sin i = 13.6$\kms\
as a more plausible detection threshold for their data, and that is
what is used here. With similar reasoning only values of $v \sin i \geq
10$\kms\ are used from the work of Sicilia-Aguilar et al.  (2005). The
sample is also filtered of poor quality data with a fractional $v \sin
i$ uncertainty greater than 25 per cent to prevent unnecessarily
broadening the observed $\sin i$ distribution.

Stellar radii are estimated from combining $L$ and $T_{\rm eff}$
(Hillenbrand 1997). $L$ comes from extinction-corrected $I$-band magnitudes
and bolometric corrections, and assumes a distance to the ONC of
470\,pc. $T_{\rm eff}$ values are estimated from the relationship
between spectral type and $T_{\rm eff}$ (with some small modifications)
proposed by Cohen \& Kuhi (1979). Hillenbrand (1997) estimates that $L$
is uncertain by about 0.2\,dex due to variability and uncertainties in
extinction. Random errors in spectral types lead to uncertainties of
about 0.02\,dex in the $T_{\rm eff}$ values. As $R \propto L^{-1/2}
T_{\rm eff}^2$, the random errors in $R$ are about 0.11\,dex. It is
fair to say that this uncertainty is itself uncertain, and so values
between 0.07\,dex and 0.15\,dex are tested in the modelling.  Hillenbrand
(1997) also had concerns that because of difficulties in assigning an
extinction value, the $L$ and hence $R$ of accreting classical T-Tauri
stars (CTTS) would be {\em under}estimated.  As the rotational database
contains information on the $I-K$ excess and EW[Ca], both of which are 
diagnostic of strong accretion, such stars can be optionally excluded.

The final database contains 95 objects with both adequate $P$ and $v\sin i$
measurements and estimates of $L$, $T_{\rm eff}$ and $R$.
These objects are listed in Table~\ref{database}
(available in electronic form only) and the HR diagram is compared with
the stellar evolutionary models of Siess, Dufour \& Forestini in
Fig.~\ref{hrdiag} (2000 --
the variant with a metallicity of 0.02 and no convective overshoot).

\subsection{Selection effects and biases}

The database is subject to a number of selection effects and biases
(see also the discussion in Herbst \& Mundt 2005).  There is a bias
against the inclusion of CTTS and in favour of weak-lined T-Tauri stars
(WTTS). This is because CTTS often show accretion-related, non-periodic
variability which masks the true rotation period of the star. There is
also a bias towards objects with shorter rotation periods simply
because these are easier to measure from datasets with a limited time
span. Neither of these selection effects will bias the observed
$\sin i$ distribution unless CTTS and more slowly spinning stars have
non-random spin-axis orientation.

Of more concern is the observational bias against slow rotators and
objects with small $\sin i$ because of the limited sensitivity of the
$v \sin i$ measurements. In principle this can be accounted for in our
modelling by putting a $v \sin i$ cut-off in the model. This threshold
imposes a smooth roll-off in the values of $\sin i$ that are capable of
detection. The exact shape of the roll-off depends on the distribution
of true equatorial velocities (see section~\ref{modelvtrue}).

A further consideration is the bias away from low inclination systems
due either to the lack of visibility of starspots at equatorial
latitudes caused by limb darkening, or to the reduced amplitude of spot
modulation caused by starspots at higher latitudes (see discussion in
O'Dell \& Hendry 1994). The actual latitude distribution of spots on
very young stars is still debatable. There have been theoretical
predictions of polar concentrations for spots on rapidly rotating stars
with deep convection zones (e.g. Sch\"ussler et al. 1996). However,
Granzer et al. (2000) predict spots over a wide range of latitude for
fast-rotating T-Tauri stars, with an equatorial concentration in slower
rotators. The evidence from Doppler imaging of spots on very young
T-Tauri stars is mixed.  Some show spot activity at low latitudes, some
at high latitudes and other have spots at all latitudes (see Joncour,
Bertout \& Bouvier 1994; Johns-Krull \& Hatzes 1997; Neuh\"auser et
al. 1998).  For now I make the assumption that there is some
inclination $i_{\rm th}$, below which starspot modulation is never seen
(see also O'Dell et al. 1994). This threshold is allowed to vary over
some plausible range or can be tuned to give the best fit to the
lower end of the observed $\sin i$ distribution.

\section{Modelling the observed inclination distribution}

\subsection{The general approach}

The observational estimates of $\sin i$ for each star are given by
\be
(\sin i)_{\rm obs} = \left(\frac{k}{2\pi}\right)\,\frac{P_{\rm obs}\, (v \sin
i)_{\rm obs}}{R_{\rm obs}}\, ,
\label{eqn1}
\ee
where $P_{\rm obs}$ is the observed period, $(v\sin i)_{\rm obs}$ is
the observed projected equatorial velocity, $R_{\rm obs}$ is the
estimate of the stellar radius based on a distance to the ONC of
470\,pc, and $k$ is a constant which depends on the units used for the
various quantities. The aim is to model the distribution of $(\sin
i)_{\rm obs}$ with a Monte Carlo simulation. To that end I define
\be
k = 2\pi\, \frac{R_{\rm true}}{P_{\rm true}\, v_{\rm true}}\,
\label{eqn2}
\ee
where the subscript ``true'' indicates the actual values of these three
parameters in the absence of measurement uncertainties and other
systematic effects (see below). I assume that 
$P_{\rm obs}$ is related to $P_{\rm true}$ by
\be
P_{\rm obs} = P_{\rm true}\, ( 1 + \delta_P U)\, ,
\ee
where $\delta_P$ is the fractional uncertainty in the period and $U$ is
a random number drawn from a Gaussian distribution with a mean of zero
and unit standard deviation. Similarly,
\be
(v \sin i)_{\rm obs} = v_{\rm true}\, \sin i\, (1 + \delta_v U)\,
\label{eqn4}
\ee
where $\sin i$ is drawn randomly assuming that $\cos i$ is distributed
uniformly between 0 and 1 (i.e. random orientation of the spin
axes) and $\delta_v$ is the fractional uncertainty
in the $v \sin i$ measurements. The true velocity and $\sin i$ are
split into separate factors because the effect of an 
observational lower limit to $v \sin i$ can only be modelled properly if a
distribution of $v_{\rm true}$ is initially specified. For instance a
star with $v_{\rm true}=50$\kms\ cannot be in the Rhode et al. (2001) 
observational sample unless $\sin i > 13.6/50$, but clearly this
threshold changes with $v_{\rm true}$ (see section~\ref{modelvtrue}).

There is a similar expression relating $R_{\rm obs}$ and $R_{\rm true}$ 
but this must also take into account that the true distance, $D$, which
is assumed common to all stars in the sample, may 
differ from the 470\,pc assumed by Hillenbrand (1997). In addition a
term must be included that admits the possibility that a given star
could be an unresolved binary system. In which case, $L$ is the sum of two
components and  $R_{\rm true}$
could be smaller than $R_{\rm obs}$ by a factor, $1\leq b\leq\sqrt{2}$,
for binary mass ratios $0\leq q\leq 1$. In this case
\be
R_{\rm obs} = R_{\rm true}\,\left(\frac{470\,{\rm
    pc}}{D}\right)\,b(q)\,10^{\delta_{\log R} U}\, ,
\label{eqn5}
\ee
where $\delta_{\log R}$ is the uncertainty (in dex) of the logarithmic
radius estimate (see section 3.1).

Hence, combining equations 1--5 the model for $(\sin i)_{\rm obs}$ can
be expressed as
\be
(\sin i)_{\rm obs} = \left[\frac{D}{470\,{\rm pc}}\right]\,\left[\frac{(1+\delta_P U_1)\,(1+\delta_v
    U_2)}{b(q)\,10^{\delta_{\log R} U_3}}\right]\, \sin i\, ,
\label{eqn6}
\ee
where $U_1$, $U_2$, $U_3$ indicate that these are three 
  different random numbers taken from a unit Gaussian distribution.

To estimate the distance to the cluster I adopt the simplest, unbiased
approach suggested by Hendry et al. (1993). If equation~\ref{eqn6} is
  re-expressed as
\be
(\sin i)_{\rm obs} = \left[\frac{D}{470\,{\rm pc}}\right]\, \alpha\, ,
\label{eqn7}
\ee
then taking the average over all the stars in the sample and the
average over all the trials in a model Monte-Carlo simulation, the best estimate for the
distance to the cluster is
\be
D = 470\, \left(\frac{\langle (\sin i)_{\rm
    obs} \rangle }{\langle\alpha \rangle}\right)\,{\rm pc}\, .
\label{alpha}
\ee

\subsection{The true equatorial velocity distribution}
\label{modelvtrue}

The model distribution for $v_{\rm true}$ has some influence on the
derived cluster distance. The reason is that the lower
threshold for $v\sin i$ measurements leads to a higher threshold in 
$(\sin i)_{\rm obs}$ for lower values of $v_{\rm true}$.
To obtain a good model for the distribution of $(\sin i)_{\rm obs}$
requires a reasonable description of the distribution of $v_{\rm
  true}$. An incorrect $v_{\rm true}$ distribution leads to a
systematic error in the distance estimate.

With the size of sample considered here this error is not entirely
negligible compared with the statistical uncertainties in $D$ (see
section~\ref{results2}).  Fortunately we can check that the assumed
$v_{\rm true}$ distribution is reasonable by comparing the modelled
distribution of $v\sin i$ from equation~\ref{eqn4} (after uncertainties
and selection thresholds have been applied) with that seen in the
data. For instance a flat $v_{\rm true}$ distribution would not explain
the observed $v \sin i$ distribution in the ONC (see
section~\ref{results2}).

\subsection{The binary correction factor}
\label{binary}

\begin{figure}
\includegraphics[width=75mm]{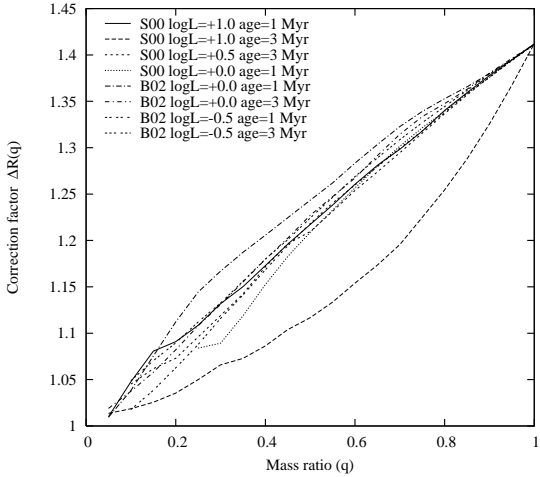}
\caption{The multiplicative binary correction factor as a function of
  mass ratio for ages of 1\,Myr or 3\,Myr, for
$-0.5<\log (L/L_{\odot})<1.0$ and for two different evolutionary models
  (Siess et al. 2000 -- S00 and Baraffe et al. 2002 -- B02).
}
\label{rcorr}
\end{figure}

Unresolved binary systems will have systematically overestimated $L$,
overestimated $R$ and hence underestimated $\sin i$. A neglect of this
effect would lead to an underestimated distance in
equation~\ref{alpha}.  In older clusters where stars have reached the
ZAMS, unresolved binaries with $q\ga 0.5$ are easily detected as they
lie significantly above the cluster single stars locus in the HR
diagram. This becomes impractical in young clusters like the ONC where
there may be an age spread that is a significant fraction of the mean
age, and where differential extinction and variability cause
significant spreads in any CMD.

I have calculated a correction to $R$ that should be applied 
to an unresolved binary system of total luminosity $L$ and
mass ratio $q$. The calculation uses a model of $L$ versus $T_{\rm
  eff}$ calculated at a range of masses. For this paper I have
used the models of Siess et al. (2000) and 
Baraffe et al. (2002 -- the variant with a mixing length equal
to the pressure scale height). For an assumed age and $L$, an
initial guess (an upper limit) 
for $m_1$ is taken straight from the isochrone and an
initial guess for $m_2$ is $q m_1$. The value of $m_1$ is 
iteratively reduced until the total luminosity of the two components
equals $L$. At this point the actual $T_{\rm eff}$ of the primary and
secondary can be calculated as well as a flux-averaged $T_{\rm eff}$
that is assumed to correspond to the observed spectral type.
The multiplicative overestimate of the radius is given by
\be
\Delta R(q) = \left(\frac{L}{L_1}\right)^{1/2}\,\left(\frac{T_{\rm
    eff,1}}{T_{\rm eff}}\right)^{2}
\label{deltar}
\ee

In Fig.~\ref{rcorr} I show $\Delta R(q)$ for $-0.5<\log L/L_{\odot}<1$
and for ages of 1 or 3\,Myr that cover the majority of stars in the ONC
rotation sample. At low luminosities the Baraffe et al. (2002)
models are used because the Siess et al. (2000) models do not extend to low
masses ($M<0.1\,M_{\odot}$), on the other hand the Baraffe et
al. models do not extend to high luminosities as they are limited to
masses less than $1.4\,M_{\odot}$. At luminosities where the models
overlap there is good agreement.  These curves demonstrate that $\Delta
R(q)$ is quite insensitive to changes in age, but does behave
differently for older stars with high luminosity. This is caused by a sharply
changing slope in the luminosity-mass relationship as stars develop
a radiative core. This occurs at about $10\,L_{\odot}$ at 3\,Myr, but
at $30\,L_{\odot}$ for ages of $\leq 1$\,Myr. This boundary
more-or-less coincides with the envelope of the sample stars considered
here (see Fig.~\ref{hrdiag}) and so a single $\Delta R(q)$
curve was adopted that is representative for most of the sample
(the Siess et al. model with $\log L/L_{\odot} = 0.5$ at 3\,Myr),
which is given the label ``r1'' in section~\ref{results3}. To check the
sensitivity of the results to this assumption the most deviant $\Delta
R(q)$ curve in Fig.~\ref{rcorr} 
(the Siess et al. model with $\log L/L_{\odot} = 1.0$ and
at 3\,Myr) is also tested in section~\ref{results3}
and given the label ``r2''. The difference in derived distance between
using models r1 and r2 turns out to be small (2 per cent), but 
the neglect of binaries altogether would lead to a distance
underestimate of $\simeq 10$ per cent.

The relationship between $\Delta R(q)$ and $b(q)$ in
equation~\ref{eqn6} is set by assuming that a fraction $f$ of
star-systems are binaries, with $q$ drawn randomly from a defined
distribution. The appropriate value of $f$ and the $q$ distribution can
be inferred from observational studies of the ONC.

Using different high spatial resolution surveys Prosser et al. (1994),
Padgett, Strom \& Ghez (1997), Petr et al. (1998), Simon, Close \& Beck
(1999) and K\"ohler et al. (2006) have concluded that the binary
frequency in the range 100--1000\,au is similar to that of field stars
and probably declines towards lower masses in a similar way. Details
for closer binary systems are yet to be published, but preliminary
results suggest that the close binary frequency is also similar to that
of field stars (Stassun \& Mathieu 2006).

I make the assumption that binary properties in the ONC are
similar to that of the field. Duquennoy \& Mayor (1991) show that the
overall binary frequency for solar-mass stars is 57 per cent and at
distances of $\sim 450$\,pc, the vast majority of these would be
unresolved binaries in the ONC sample. The binary frequency declines
towards 30--40 per cent in M dwarfs with 0.3--0.5\,$M_{\odot}$ (Fischer
\& Marcy 1992). The mass ratio distribution appears reasonably uniform,
at least over the range $0.5<q\leq 1$ that is most important for the
phenomena considered here.  The stars in our rotational sample cover a mass range
of $0.2<M/M_{\odot}<2.5$ according to the PMS tracks of Siess et
al. (2000; see Fig.~\ref{hrdiag}).  Rather than complicate things
further,  the binary frequency in the ONC is assumed to be 50 per cent,
mass-independendent and  $q$ is given a uniform distribution. The
sensitivity of the results to these assumptions is tested
by allowing the binary frequency to vary by $\pm 20$ per
cent in the simulations (see section~\ref{results3}).

It is worth stressing that one could just use an average value of
$b(q)$ as a multiplicative correction factor to the distance estimate
in equation~\ref{alpha}. However, I calculate individual $b$
values for each star in the Monte Carlo simulations because binarity
introduces an extra spread in the model $\alpha$
distribution which should be taken into account when deciding whether
the model is a reasonable fit to the data and, for instance, in
deciding on an appropriate value for $i_{\rm th}$ (see section 3.2).

As a final remark on binary systems, I note that the $v \sin i$
measurements of very short period (less than a few days) binaries could
be artificially increased by orbital motion because of the lengthy
exposure times used to obtain the spectra (up to a few hours for the
Rhode et al. [2001] measurements). This would increase $(\sin i)_{\rm
obs}$ for the affected objects, leading to a distance overestimate.  I
have neglected this possibility in the modelling, but as only a very
small minority of binary systems are likely to have such short periods
and be observed at unfavourable phases, this seems reasonable.

\subsection{Recap of assumptions and the step-by-step method}

Having described the general methods I use to calculate a model $(\sin i)_{\rm
  obs}$ distribution, the following summarises the specific assumptions 
and steps in the modelling process.

\begin{enumerate}

\item Choose the sample. Stars are selected from the rotation sample on
  the basis of having a minimum observed projected equatorial velocity
  $(v\sin i)_{\rm obs}$ (13.6\kms\ and 10\kms\ for the Rhode et
  al. [2001] and Sicilia-Aguilar [2005] observations respectively --
  see section~3.1) and can be further restricted to a range of
  temperature and whether they show signs of accretion or not (see
  section~\ref{results1}).

\item A model is chosen for the true equatorial velocity 
  ($v_{\rm true}$) distribution. Then,
  assuming that the rotation axes are randomly orientated, a set
  (typically $10^{4}$ trials per object in the sample) of model $v \sin
  i$ values are randomly generated and perturbed according to
  equation~\ref{eqn4} using the fractional uncertainties in $(v\sin
  i)_{\rm obs}$ as the $\delta_v$ values.  The model $v_{\rm true}$
  distribution is adjusted until
  a satisfactory match to the $(v\sin i)_{\rm obs}$ distribution is
  indicated by a Kolmogorov-Smirnov (K-S) test between the observed and
  modelled cumulative distribution functions (CDFs).

\item The trial values of $\sin i$ are used to calculate the $\alpha$
  values in equation~\ref{eqn7} using $\delta_v$, fractional errors in the period
  $\delta_p=0.01$, and assuming a suitable value for the fractional
  error in the $\log$ stellar radius ($\delta_{\log R}$). A fraction
  $f$ of the trials are randomly assigned a binary status and the
  appropriate correction factor to the observed radius, $b(q)$, is
  applied assuming that the binary mass ratio ($q$) is uniformly
  distributed.

\item The results of the trials pass through a filter which only allows
  those trials to proceed which have $v\sin i$ greater than the
  observational thresholds and axial inclinations greater than a
  user-defined threshold, $i_{\rm th}$.

\item At this stage the average value of $(\sin i)_{\rm obs}$ is
  divided by the average value of $\alpha$ in order to find $D$ 
  (equation~\ref{alpha}). At the same time a K-S test is
  made between the CDFs of $\alpha (D/470)$ and
  $(\sin i)_{\rm obs}$. This is used to decide whether the model is
  a reasonable description of the data (the sample is too small to
  consider $\chi^2$ tests).

\item Statistical uncertainties in the results are estimated by
  generating many fake datasets of the appropriate size. The $(\sin
  i)_{\rm obs}$ values are drawn randomly using equation~\ref{eqn7} at
  the estimated distance and these are modelled in the same way as the
  data. The standard deviation of the distance estimates is used as a
  statistical uncertainty.

\item The sensitivity of the results to the model assumptions and
  parameters are tested using different subsamples of data
  (e.g. with or without signs of accretion),
  different values of $i_{\rm th}$, $f$, $\delta_{\log R}$ or different
  models for $\Delta R(q)$ and the $v_{\rm true}$
  distribution (see section~\ref{results3}).

\end{enumerate}

\section{Results}

\label{results}
\subsection{The observed ${\bf \sin i}$ values}
\label{results1}

\begin{figure}
\includegraphics[width=80mm]{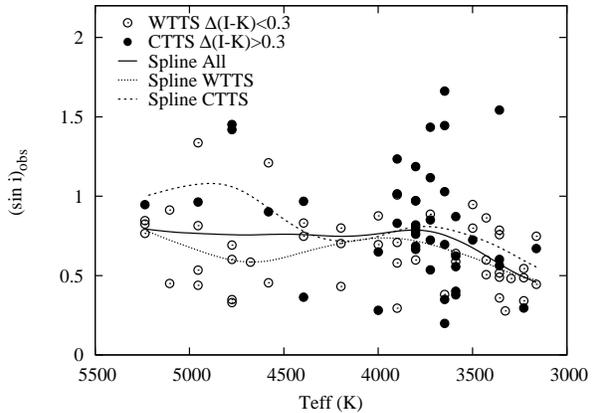}
\caption{The observed value of $\sin i$ (from equation~1) versus the
  effective temperature. CTTS/WTTS are indicated by different
  symbols. The lines show low order spline fits as a function of $T_{\rm
  eff}$ to (a) all the data, (b) the CTTS only ($\Delta(I-K)>0.3$), 
  (c) the WTTS ($\Delta(I-K)<0.3$).}
\label{siniteff}
\end{figure}

\begin{figure}
\includegraphics[width=80mm]{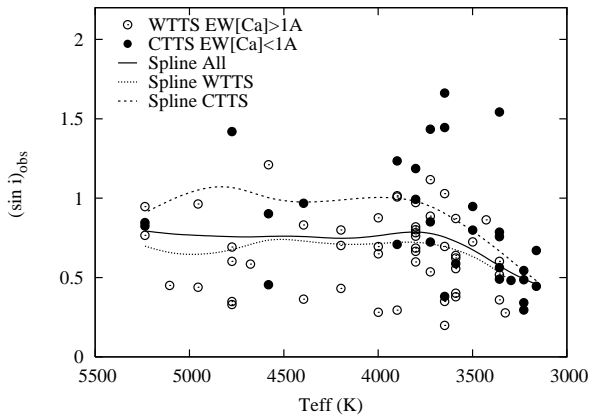}
\caption{Similar to Fig.~\ref{siniteff}, but here the WTTS and CTTS are
  classified according to whether the equivalent width of the Ca\,{\sc
  ii} 8542\AA\ line is more or less than 1\AA\ (see Hillenbrand et al. 1998).}
\label{siniteff2}
\end{figure}

In Fig.~\ref{siniteff} $(\sin i)_{\rm obs}$ versus $T_{\rm eff}$ is
plotted for the sample. The data have been distinguished on the basis
of whether their near infrared excesses, $\Delta (I-K)$, are larger or
smaller than 0.3. This division approximately separates stars with and
without active accretion and thus the two subsets are labelled as CTTS
and WTTS respectively.  Hillenbrand et al. (1998) show that because of
physical effects (disc inclination, inner holes etc) and also
observational uncertainties, this threshold is fuzzy and not capable of
establishing or excluding accretion in all cases. Unfortunately, mid-IR
measurements which are much more sensitive are only available for a
small fraction of the sample (see Rebull et al. 2006). An alternative
is to use EW[Ca], which goes into emission for strongly accreting
objects. WTTS are generally found to have EW[Ca]$>1$\AA\ and
Fig.~\ref{siniteff2} repeats Fig.~\ref{siniteff} using this criteria to
separate WTTS and CTTS.  Although crude, these diagnostics should be
sufficient to identify cases where broadband colours, magnitudes and
hence estimated luminosities and radii are likely to be affected, which
is the issue here.

Two sample selection issues are highlighted by Figs.~\ref{siniteff}
and~\ref{siniteff2}. The
first is that stars with a near infrared excess or EW[Ca]$<1$\AA\ have a
larger $(\sin i)_{\rm obs}$ on average than those without. To
illustrate this, smooth spline fits as a function of $T_{\rm eff}$ are
plotted for all the data and for the CTTS/WTTS subsets separately. The
effect is small at cool temperatures but increases
to $\simeq 30$ per cent in hotter stars. For stars with $T_{\rm eff}>3499$\,K
(see below) the average $\sin i$ is $0.78\pm 0.04$, but the averages
for the WTTS samples are $0.69\pm 0.04$ for 34 stars with $\Delta
(I-K)<0.3$ and $0.67\pm 0.04$ for 44 stars with EW[Ca]$>1$\AA.  As the
estimated distance is linearly dependent on $(\sin i)_{\rm obs}$, this
bias is of concern.  There are no physical reasons to suppose that CTTS
have larger inclination angles, but there are good reasons to suppose
that the radii of CTTS have been underestimated (leading to a $\sin i$
overestimate), either because of a systematic underestimation of the
extinction and luminosity (see Hillenbrand 1997) or because of
difficulties in determining their temperatures (see below and section
6).

The second issue is that the average $(\sin i)_{\rm obs}$ appears to
get smaller for $T_{\rm eff}<3499$\,K (corresponding to spectral types
cooler than M2) in both WTTS and CTTS.  The average $(\sin i)_{\rm
obs}$ for WTTS ($\Delta (I-K)<0.3$) with $T_{\rm eff}<3499$\,K falls
to only $0.53\pm 0.04$.  There are a number of possible causes for
this. (i) Perhaps the binary fraction is a strong function of mass (and
therefore $T_{\rm eff}$). A larger binary fraction, or tendency towards
high $q$ systems will increase the average $R_{\rm obs}$ and hence
reduce the average $(\sin i)_{\rm obs}$. The evidence from field stars
is that the binary fraction falls significantly towards lower masses,
so this explanation seems unlikely. (ii) The observed radius is derived
from $L$ and a $T_{\rm eff}$ which is derived from the Cohen
\& Kuhi (1979) main-sequence relationship between $T_{\rm eff}$ and
spectral type. For warmer stars ($T_{\rm eff}\geq 3500$\,K) this
relationship is relatively uncontroversial in the sense that
alternative scales (for instance Kenyon \& Hartmann 1995; Leggett et
al. 1996) differ by less than 2 per cent and not in a systematic
way. For cooler stars the relationship is more uncertain and especially
so for very cool PMS stars which have significantly lower gravities
than main-sequence stars of similar spectral type. The luminosity
estimates are also reliant on an accurate assessment of the extinction
which requires intrinsic broadband colours for very cool PMS
stars. Hillenbrand (1997) used the intrinsic colours of main-sequence
stars tabulated by Bessell \& Brett (1988), but very cool PMS stars may
have more giant-like colours. Dwarfs are bluer than giants by $\sim
0.3$ mag in $V-I$ for spectral types M3-M5, which would lead to a
larger extinction estimate by $\sim 0.5$ mag (at $I$).  This gives a
larger luminosity estimate by 0.2\,dex, a larger radius estimate by
0.1\,dex and hence a decreased $\sin i$ estimate by a factor of 1.26
that would almost entirely explain the trend seen.

For the reasons above the analysis here is initially restricted to the
$\sin i$ distribution of those 34 objects in Table~\ref{database} with
$\Delta (I-K)<0.3$ and $T_{\rm eff}>3499$\,K. This sample is labelled
WTTS1 in what follows. Two alternate samples are also considered,
namely 44 objects with $T_{\rm eff}>3499$\,K and EW[Ca]$>1$\AA\
(labelled WTTS2) and all 74 objects with $T_{\rm eff}>3499$\,K,
regardless of their accretion status (labelled ALL).

\subsection{The ${\bf v_{true}}$ distribution}
\label{results2}

\begin{figure*}
\centering
\begin{minipage}[t]{0.45\textwidth}
\includegraphics[width=80mm]{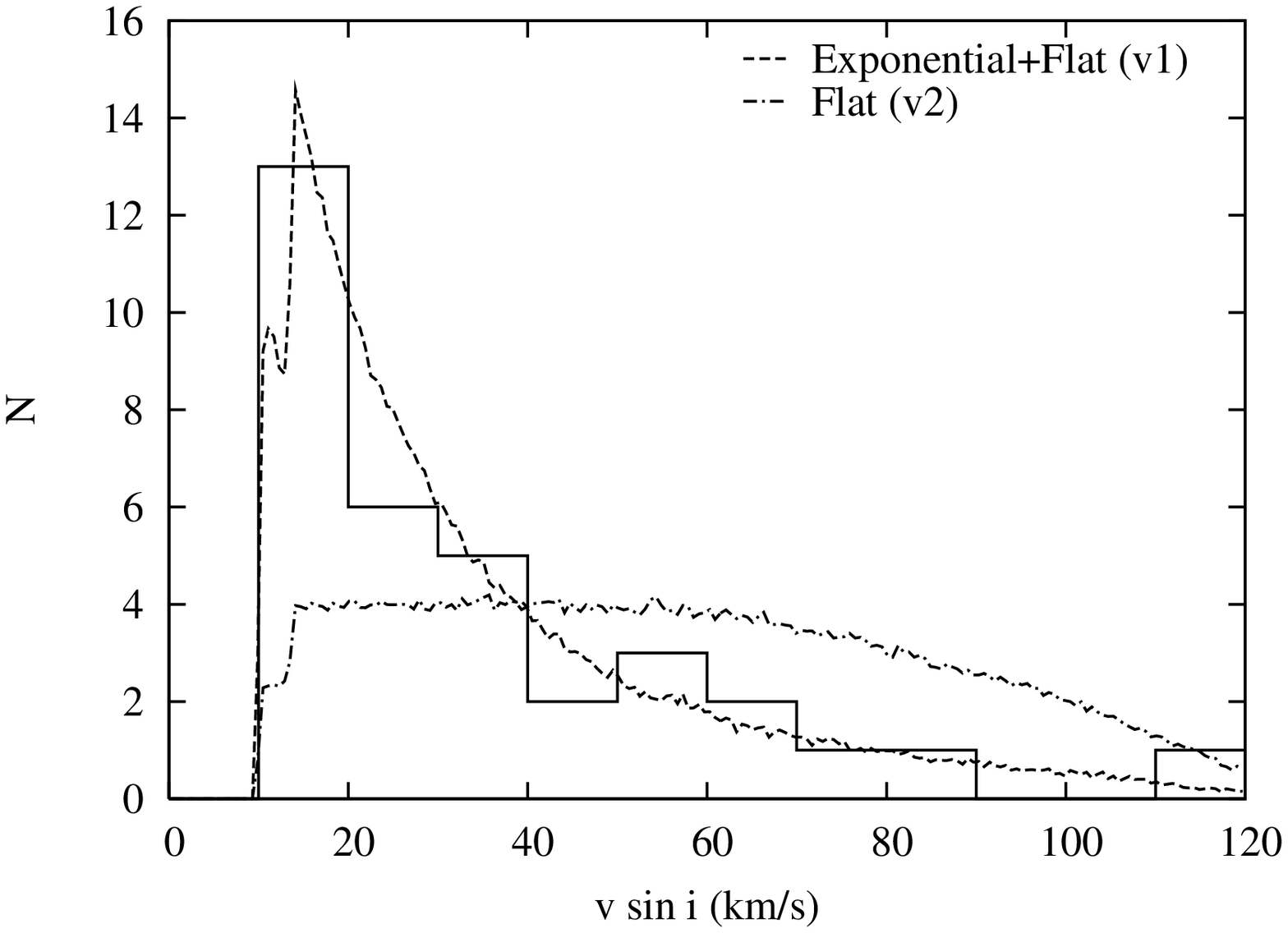}
\end{minipage}
\begin{minipage}[t]{0.45\textwidth}
\includegraphics[width=80mm]{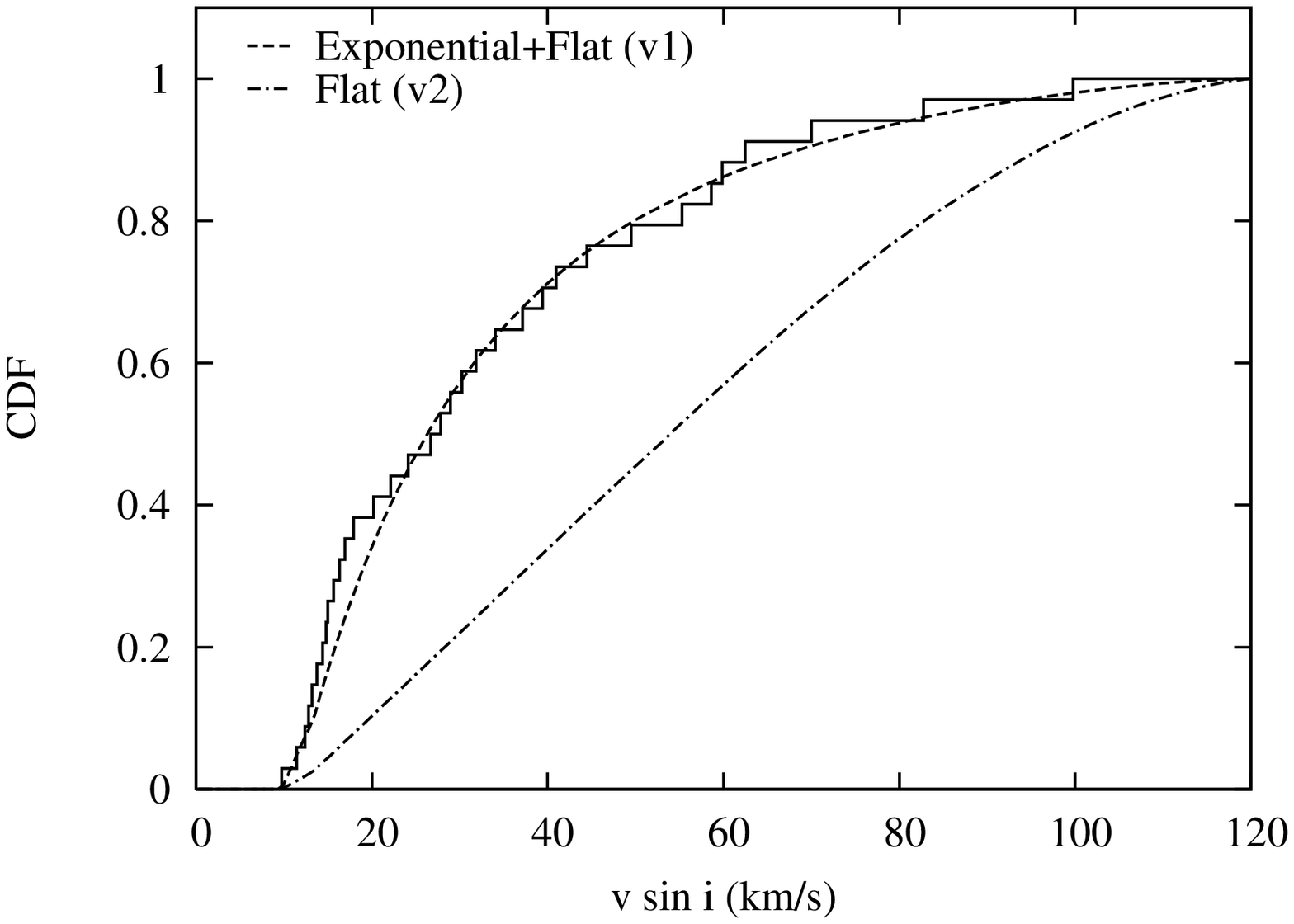}
\end{minipage}
\caption{A comparison of the observed $v \sin i$ distribution with
  Monte-Carlo simulations based on two different models of the $v_{\rm
  true}$ distributions, v1 and v2 (described in the text). The left
  panel shows the (binned) distributions and the right panel shows the
  normalised cumulative distributions used to perform a Kolmogorov-Smirnov test.
  The discontinuity in the model distributions at $v\sin i = 13.6$\kms\
  is due to the modelled sample being divided between objects with
  detectable $v\sin i$ thresholds at 13.6\kms\ and 10\kms.
}
\label{plotvtrue}
\end{figure*}

For the reasons explained in section~\ref{modelvtrue}, a reasonable
model of the $v_{\rm true}$ distribution is required and can be
constrained by comparing the simulated $v \sin i$ distribution (from
many trials) with that observed.

In Figure~\ref{plotvtrue} two of these comparisons are shown. A formal
statistical comparison has been made using a 1-dimensional K-S test of
the CDFs. The models shown have $i_{\rm th} = 30^{\circ}$, but the
model $v \sin i$ distribution does not depend strongly on this parameter.

After experimenting with a variety of simple analytical forms, 
I find that a two component $v_{\rm true}$ model (labelled v1 in
Fig.~\ref{plotvtrue}) provides a good (though not necessarily unique) 
fit to the data, with 80 per cent
of stars following an exponentially decaying distribution between
$10<v_{\rm true}<120$\kms, with a decay constant of 20\kms, and the
remaining 20 per cent being uniformly distributed between $10<v_{\rm
true}<120$\kms. The K-S test indicates a probability of 85 per cent for
the null hypothesis that the the data is drawn from the model
distribution. On the other hand, the K-S test shows that a flat
$v_{\rm true}$ distribution (labelled v2 in Fig.~\ref{plotvtrue}) can be
excluded with 99.999 per cent confidence.

\subsection{Distance estimates}
\label{results3}

\begin{figure*}
\centering
\begin{minipage}[t]{0.45\textwidth}
\includegraphics[width=84mm]{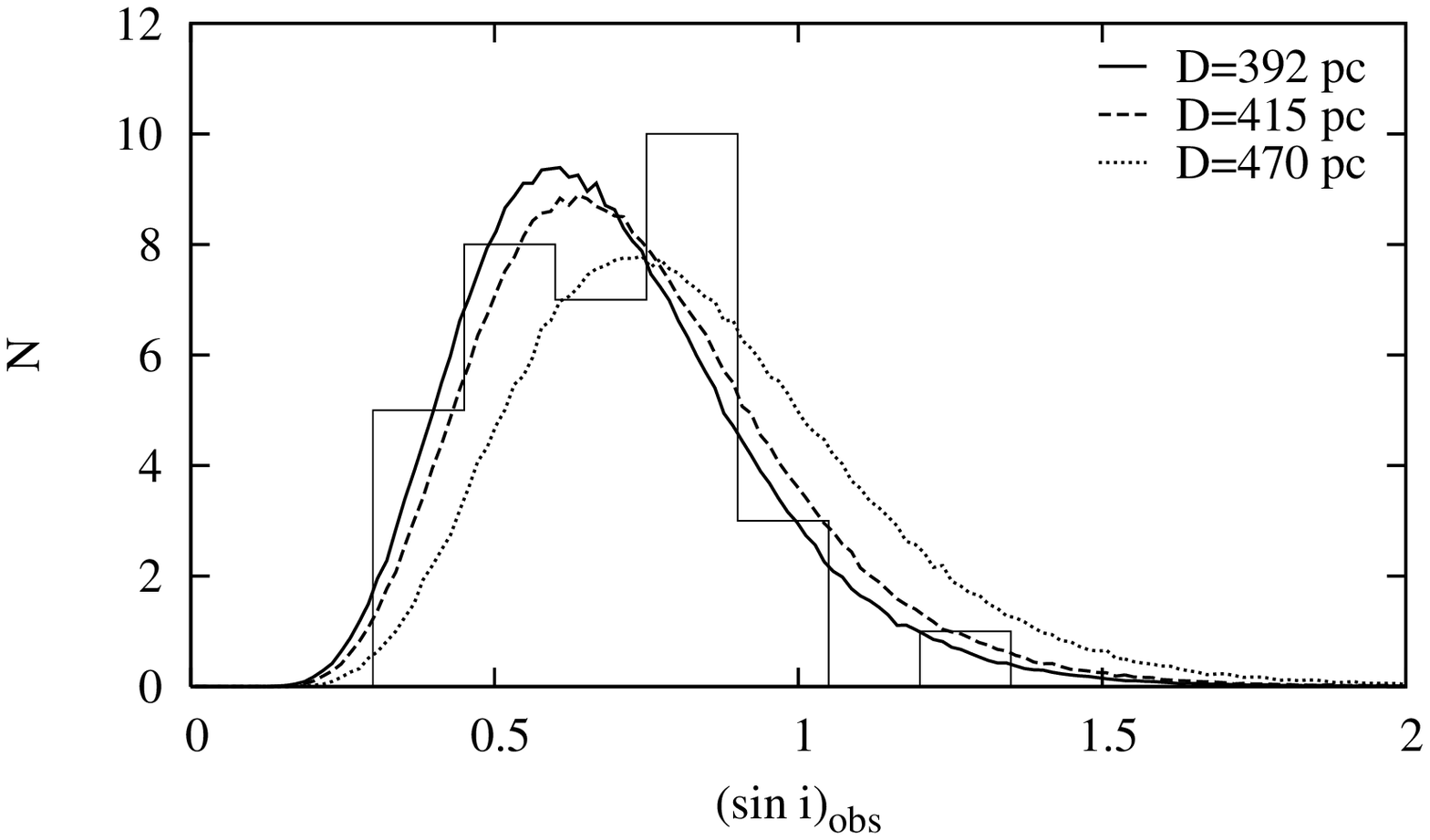}
\includegraphics[width=84mm]{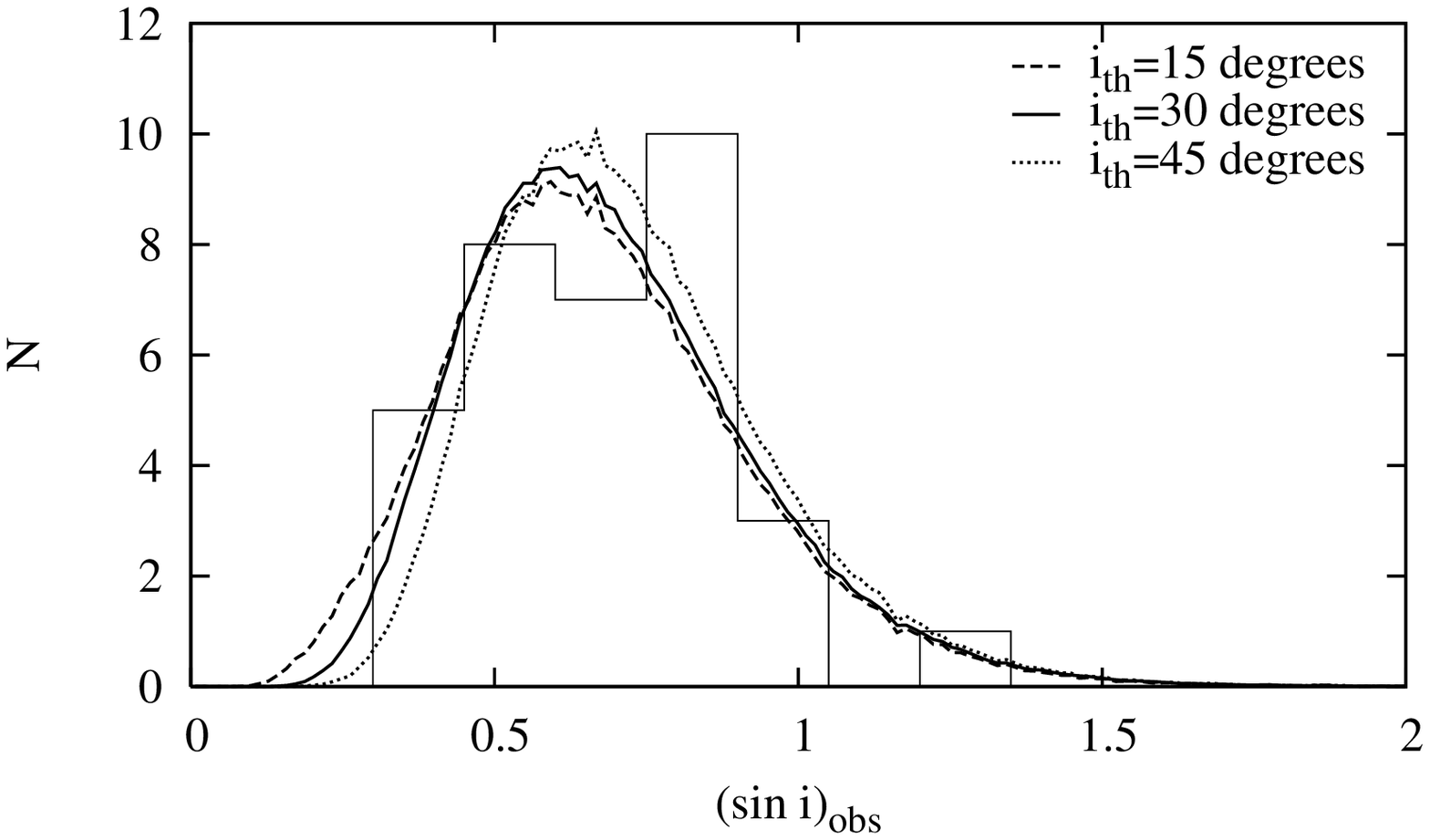}
\includegraphics[width=84mm]{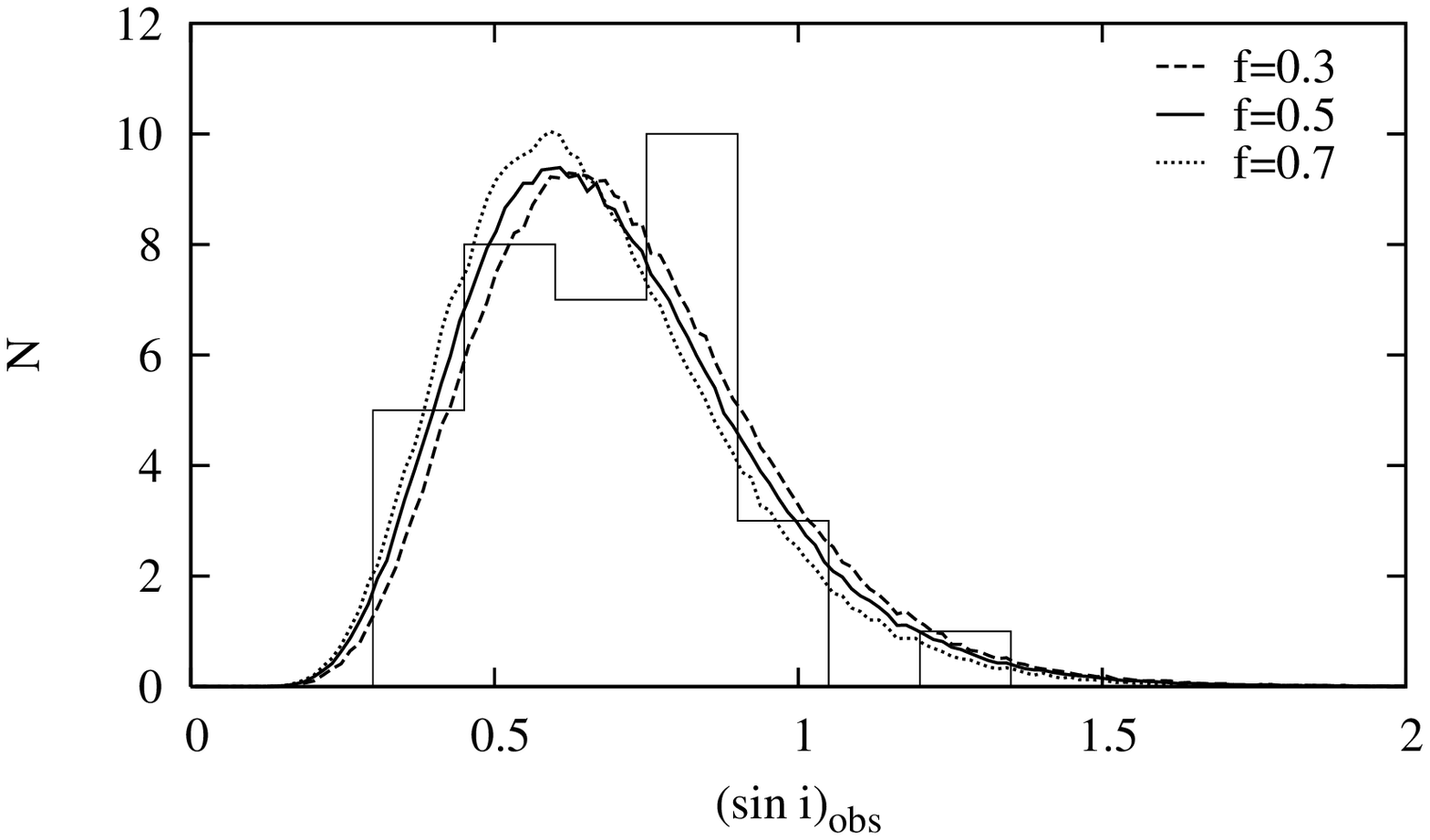}
\includegraphics[width=84mm]{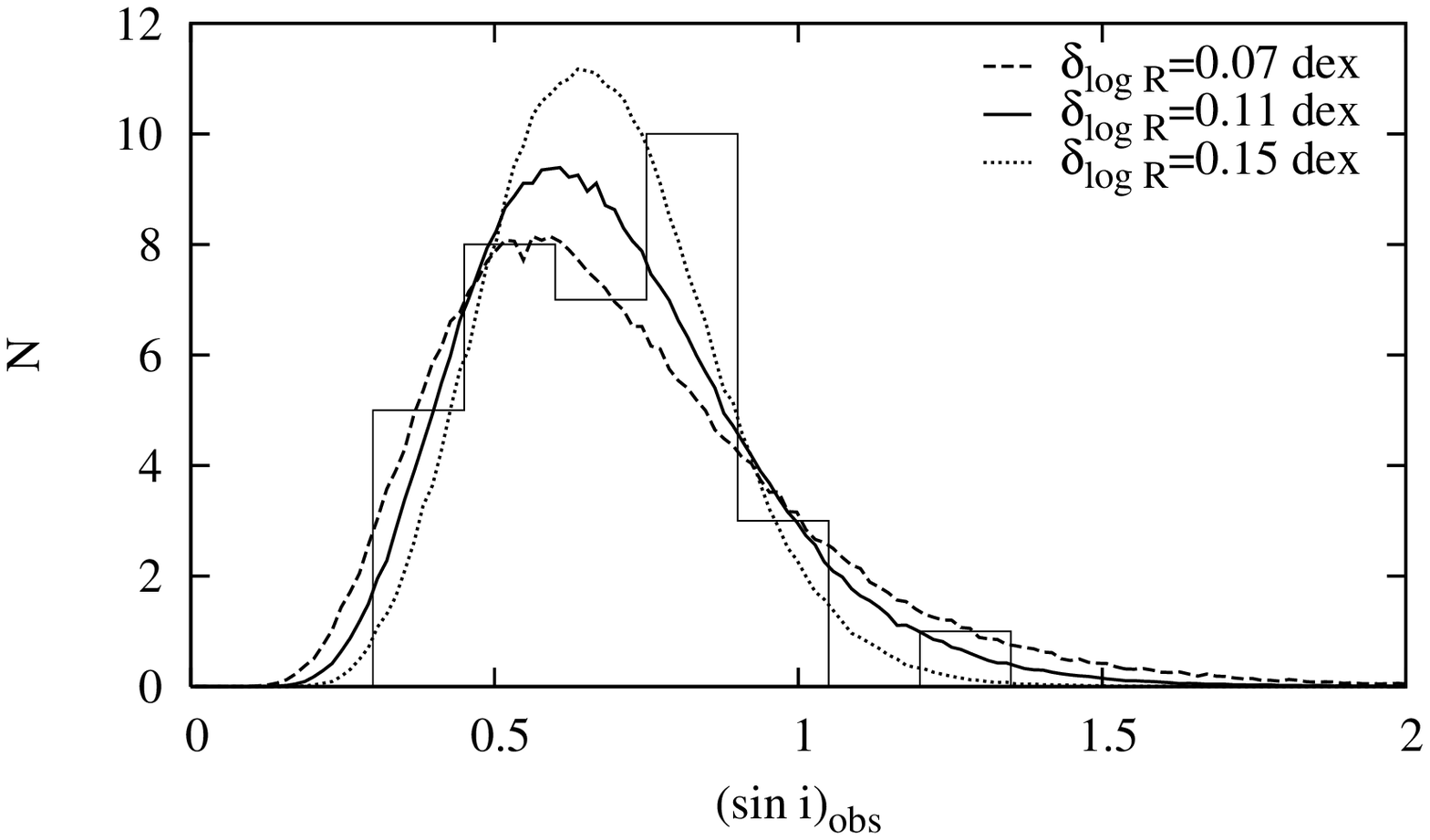}
\end{minipage}
\begin{minipage}[t]{0.45\textwidth}
\includegraphics[width=84mm]{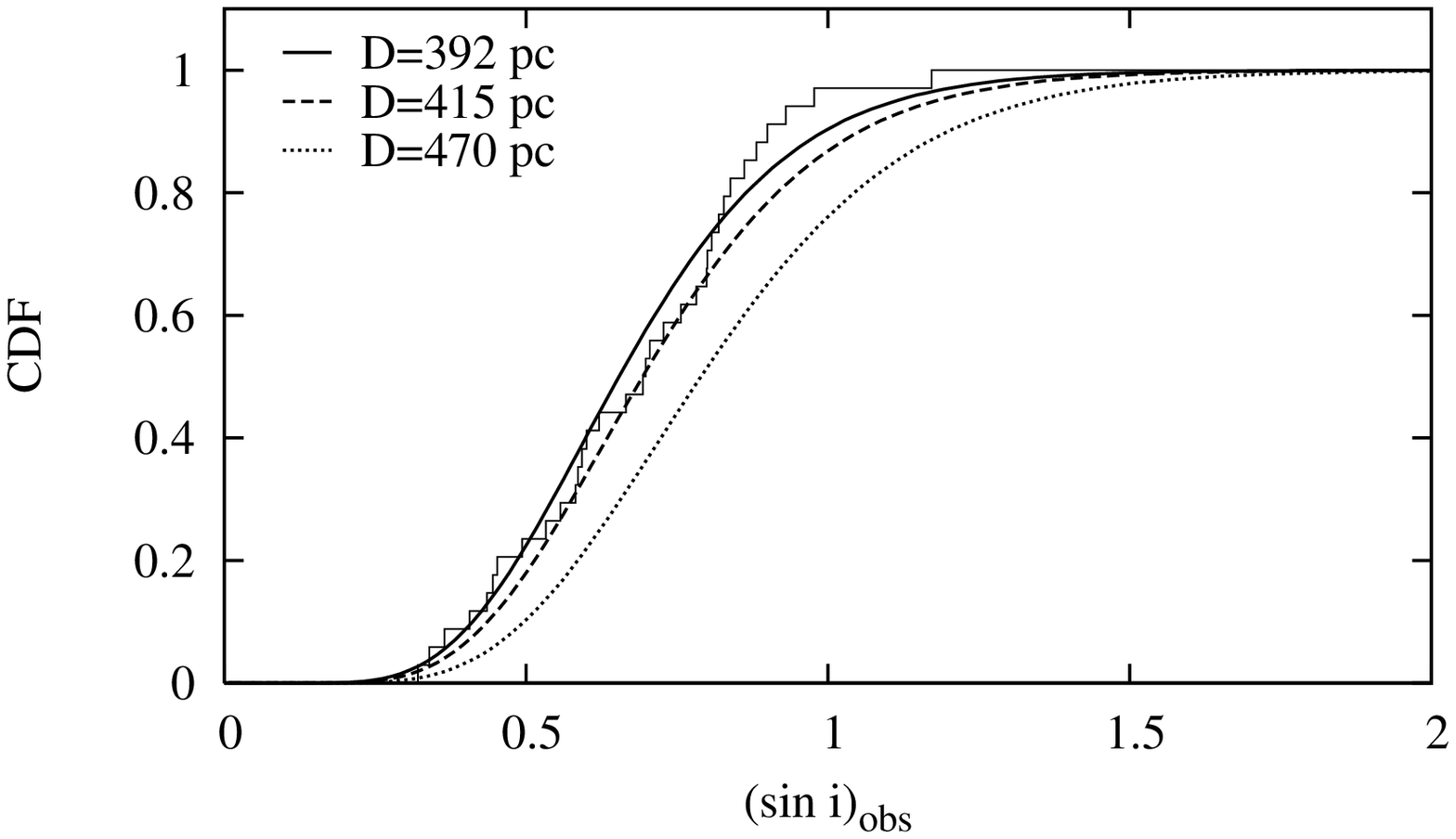}
\includegraphics[width=84mm]{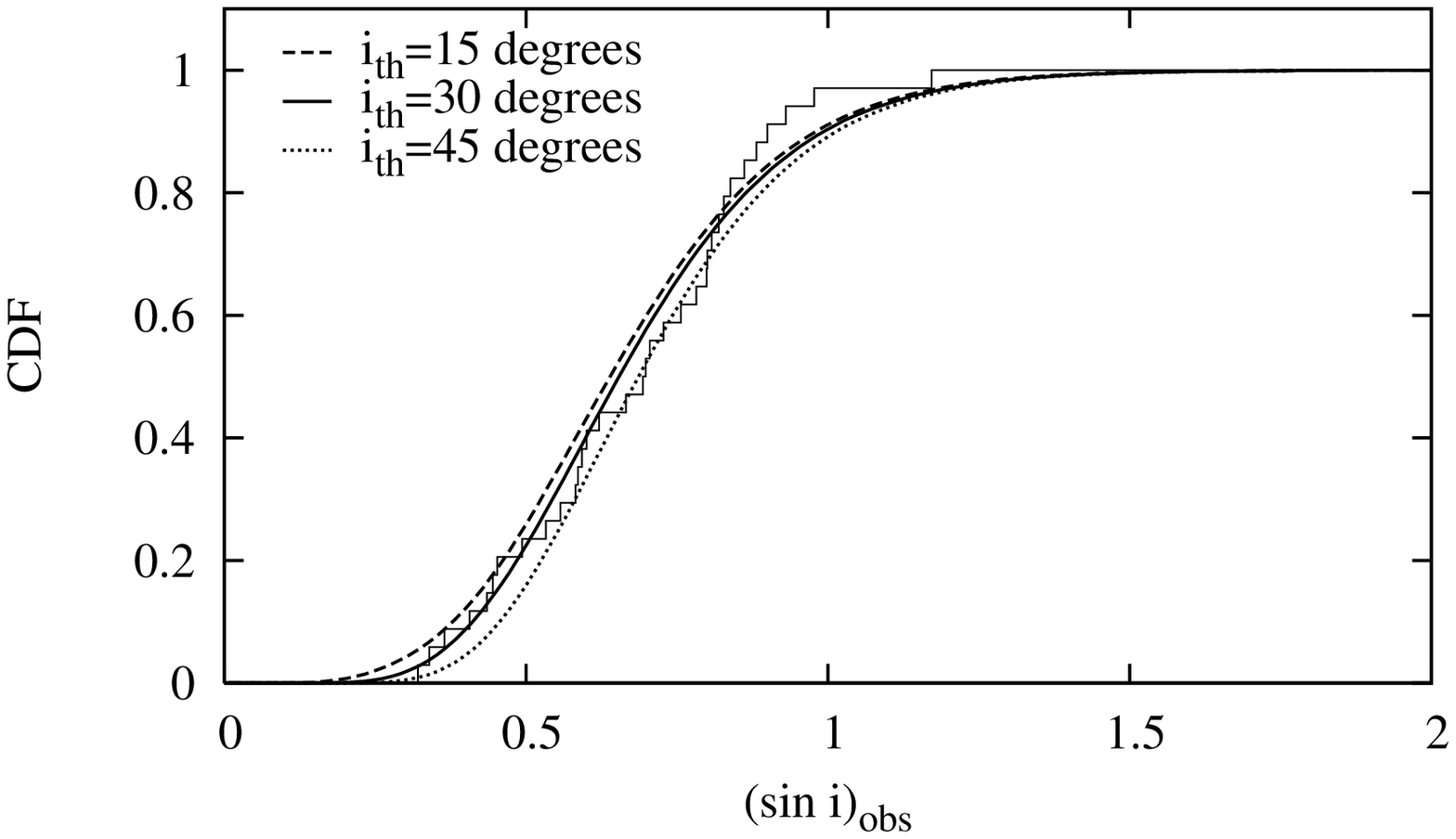}
\includegraphics[width=84mm]{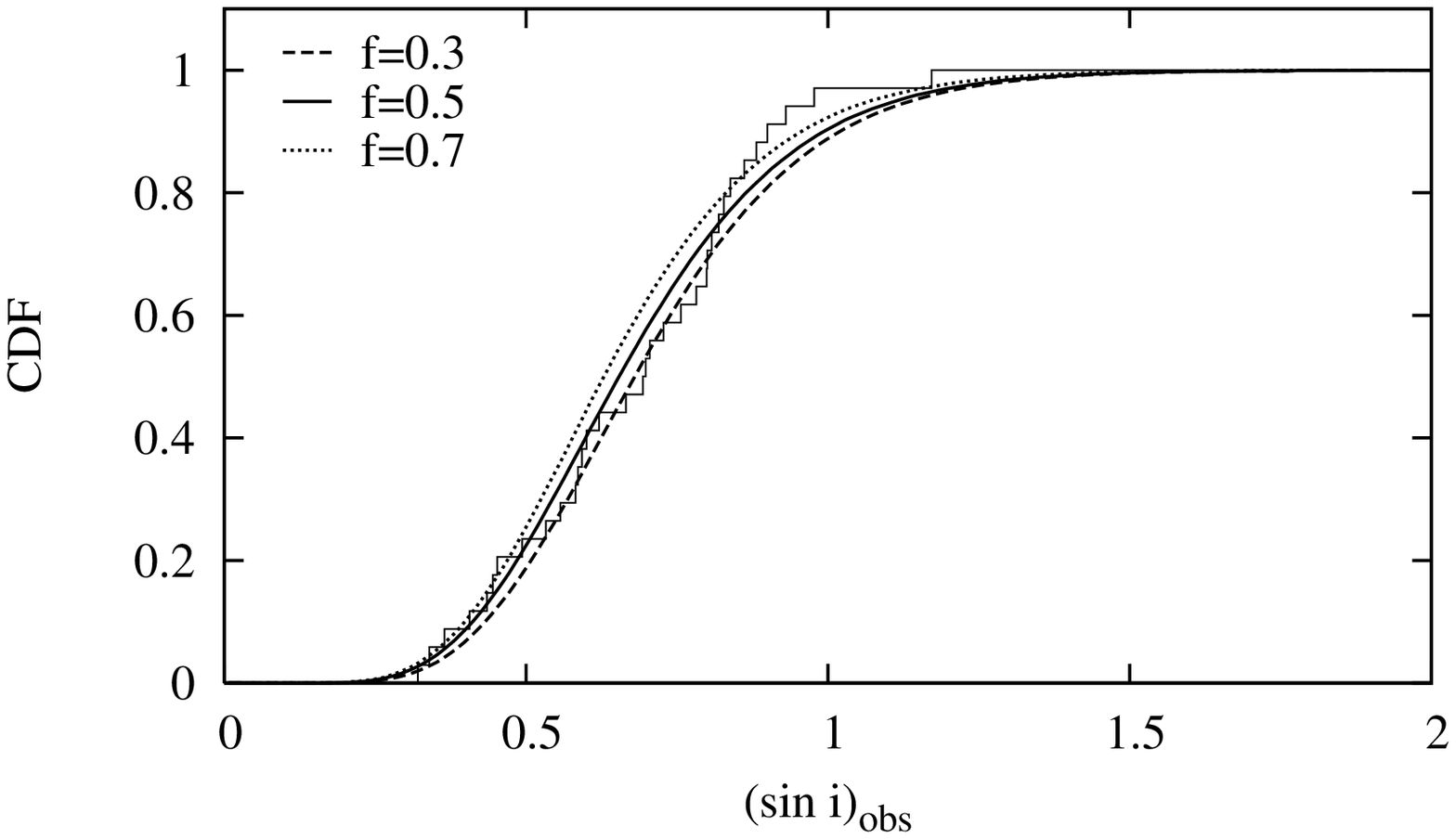}
\includegraphics[width=84mm]{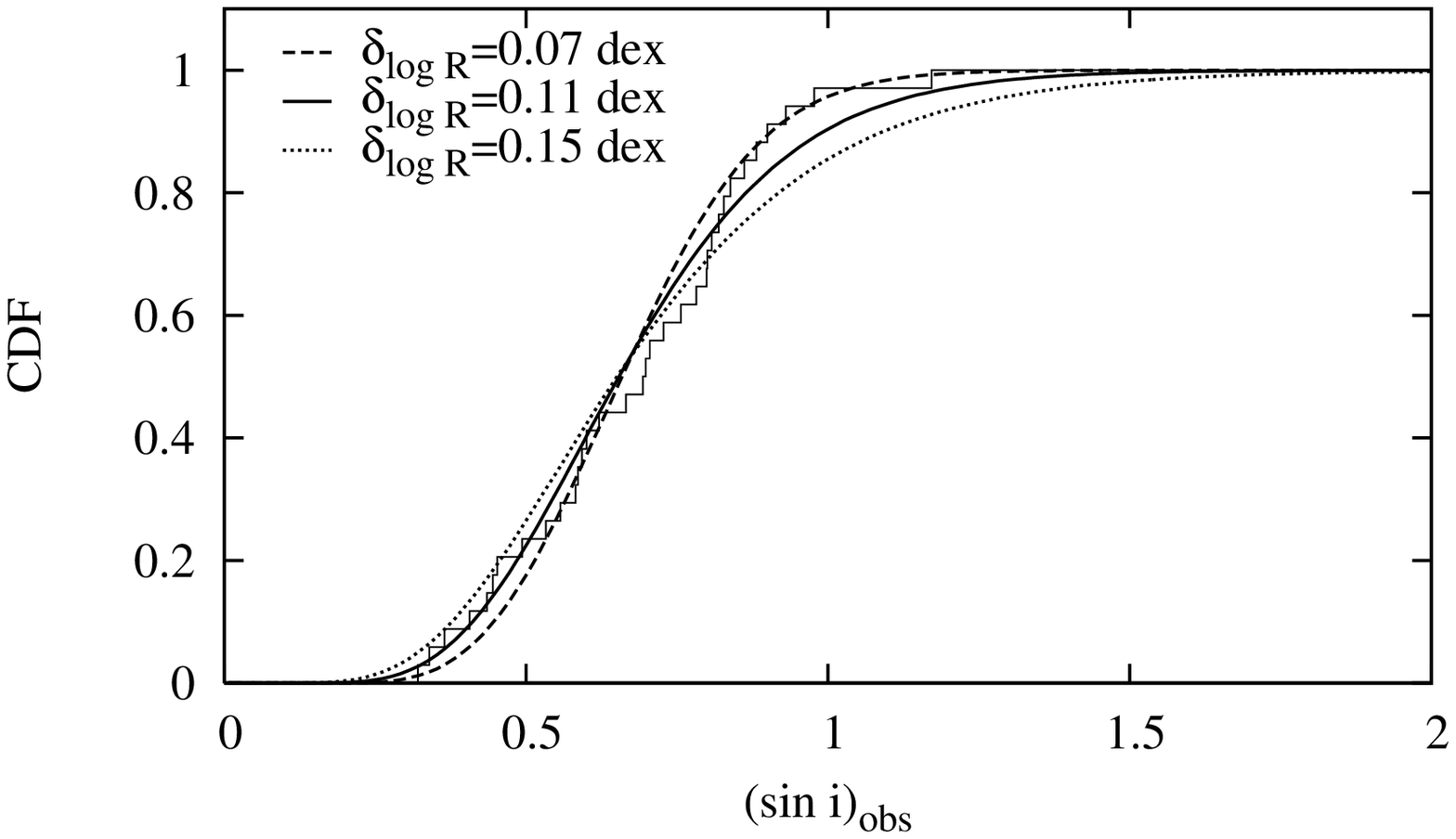}
\end{minipage}
\caption{The observed $(\sin i)_{\rm obs}$ distribution compared with
  various model distributions of the form $\alpha(D/470\ {\rm
  pc})$. The left hand panels show the differential distributions while
  the right hand panels show the normalised cumulative distributions.
(a, b) Shows the effects of distance on the data-model comparison. The
  three models shown correspons to model 1 in Table~\ref{results} at
  $D=392$\,pc, model 1 at $D=415$\,pc (corresponding to a 1-sigma
  error) and model 1 at $D=470$\,pc. (c, d) Shows the effects of
  varying $i_{\rm th}$ by displaying models 1, 2 and 3
  ($i_{\rm th}=30^{\circ}$, $15^{\circ}$ and $45^{\circ}$) at a common
  $D=392$\,pc. (e,f) Shows the effects of varying the binary fraction
  $f$ by displaying models 1, 4 and 5
  ($f=0.5$, 0.3 and 0.7) at a common $D=392$\,pc. (g, h) Shows the
  effects of varing $\delta_{\log R}$ by displaying models 1, 6 and 7
  ($\delta_{\log R}=0.11$\,dex, 0.07\,dex and 0.15\,dex) at a common
  $D=392$\,pc.}
\label{kscomp}
\end{figure*}

\setcounter{figure}{4}
\begin{figure*}
\centering
\begin{minipage}[t]{0.45\textwidth}
\includegraphics[width=84mm]{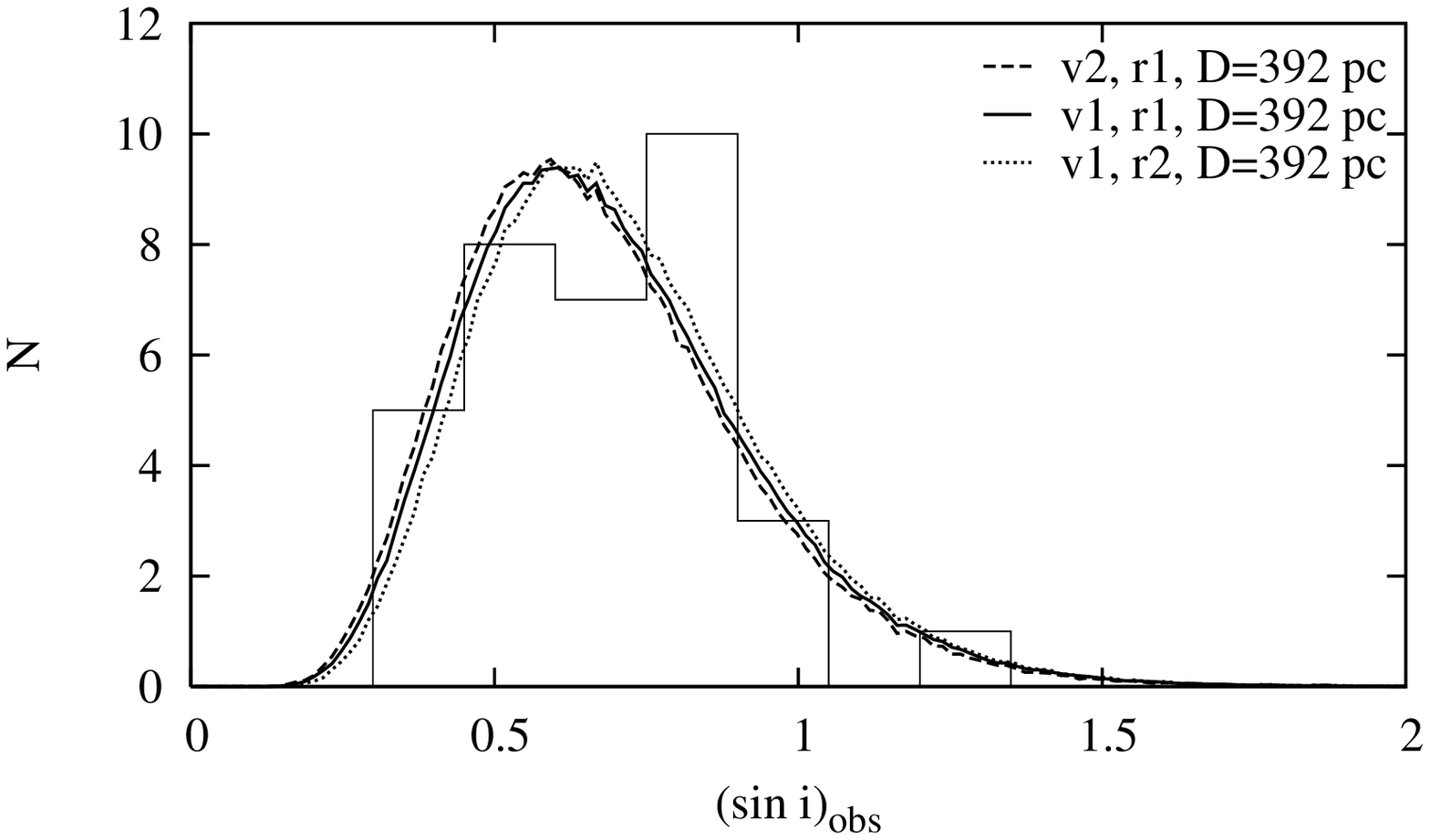}
\includegraphics[width=84mm]{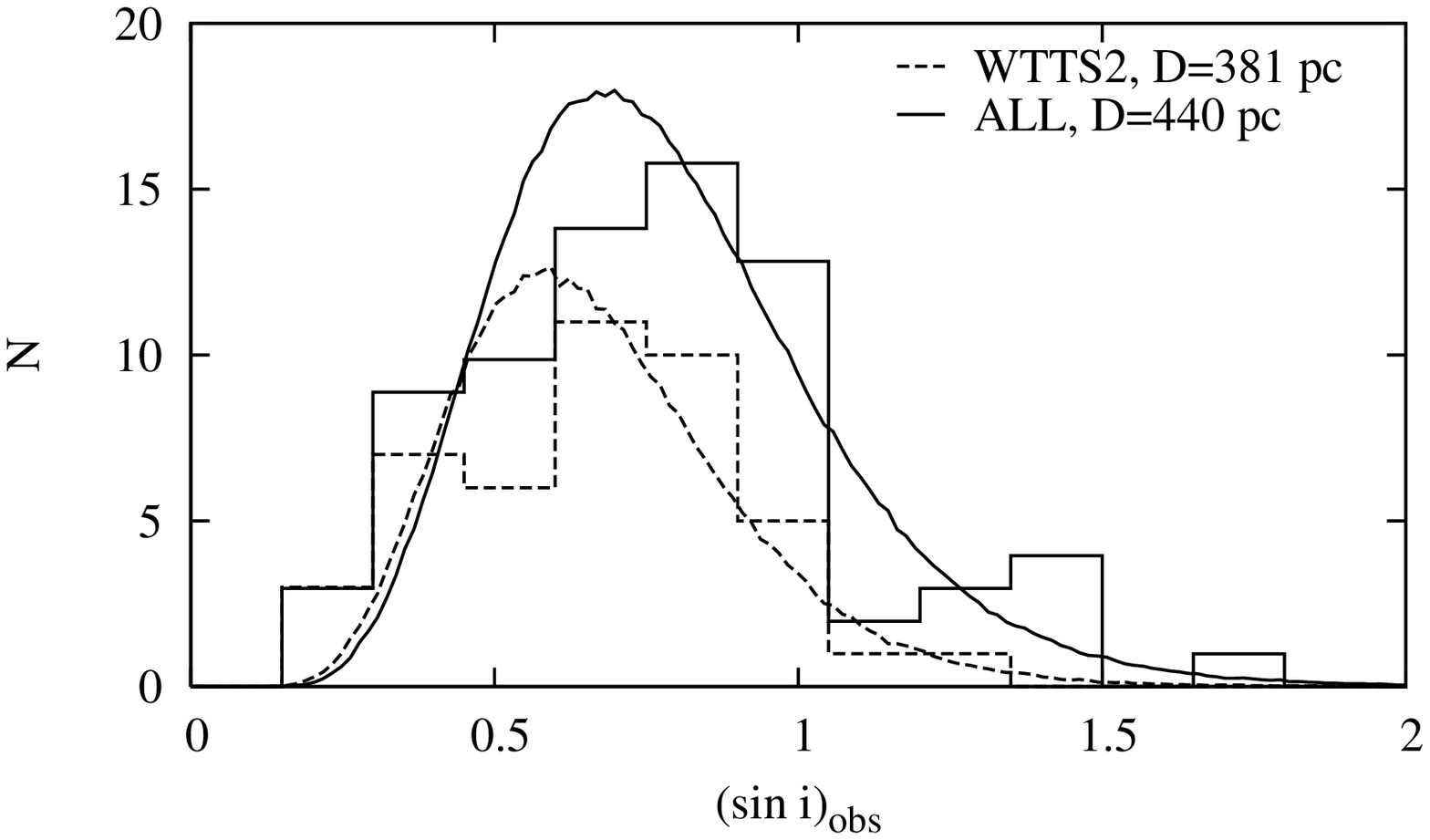}
\end{minipage}
\begin{minipage}[t]{0.45\textwidth}
\includegraphics[width=84mm]{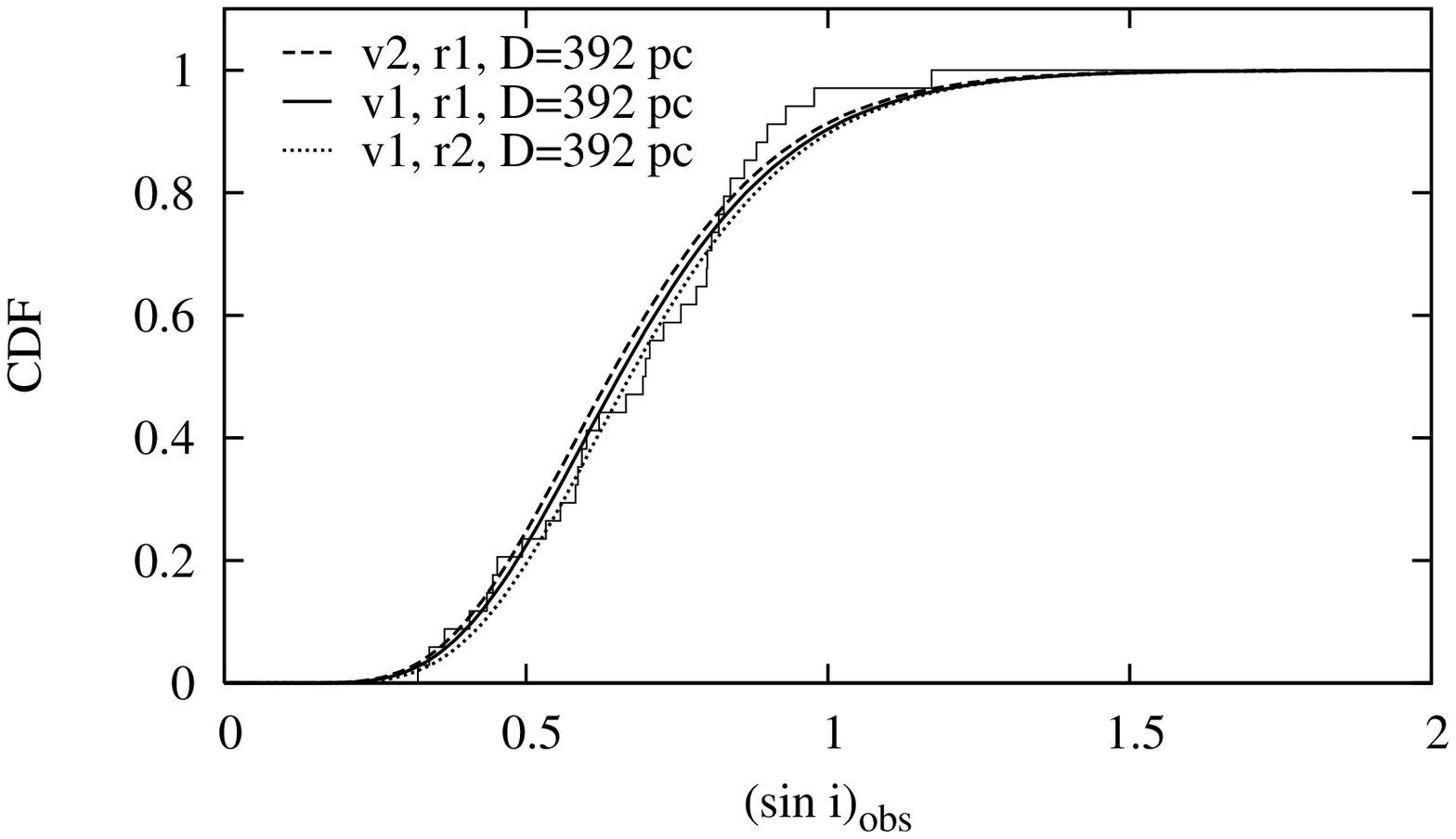}
\includegraphics[width=84mm]{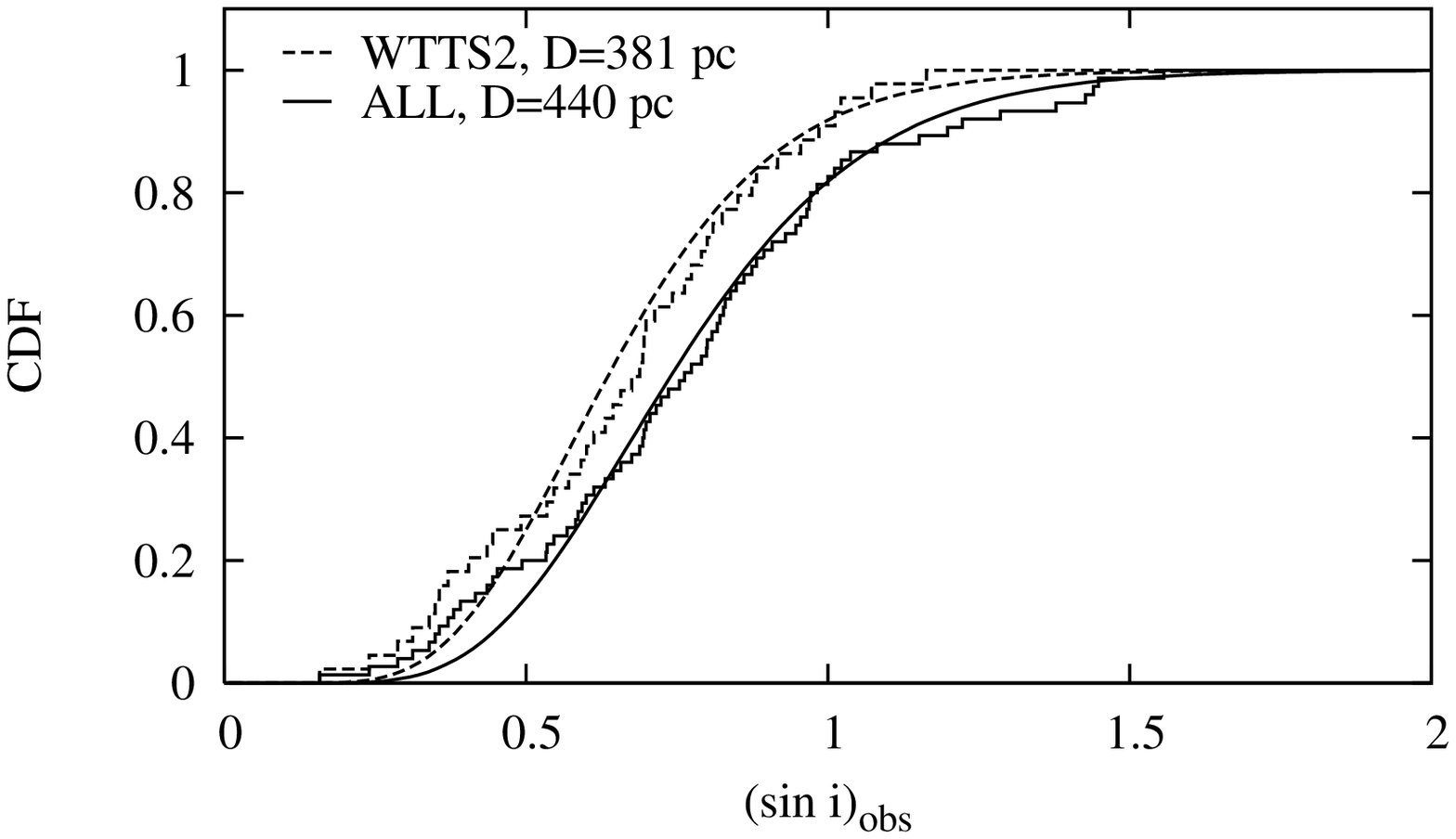}
\end{minipage}
\caption{ continued: (i, j) Shows the effects of choosing different $v_{\rm
  true}$ (v1, v2) and $\Delta R(q)$ (r1, r2) distributions by
  displaying models 1, 8 and 9 at a common $D=392$\,pc. (k, l)
  Shows a model which best fits a sample
  of WTTS defined using EW[Ca] (see section~\ref{results1})
  (model 10 at $D=381$\,pc) and a model which best fits 
  a sample containing both CTTS and
  WTTS (model 11 at $D=440$\,pc).
}
\end{figure*}

Following the procedure in section 4.4, the distributions of $\alpha$ 
and $(\sin i)_{\rm obs}$ were calculated and used to estimate the
cluster distance from equation~8. A K-S test
comparison was performed between the CDFs of $\alpha (D/470)$ and
$(\sin i)_{\rm obs}$ to test whether the model is a reasonable
representation of the data. Statistical uncertainties were estimated by
generating fake datasets of the same size, with $v_{\rm true}$ and
binary properties drawn from the distributions used in each
model. Observational errors for the fake datasets matched those in the
real dataset considered here.  The fake datasets passed through the
same analysis procedure and the standard deviations of the distance
estimates (from 300 fake datasets) were used as estimates of the
statistical error.

\begin{table*}
\caption{The results of the distance analysis. The columns are: (1) An
  analysis identification number; (2) the analysed sample and number of
  objects in the sample -- WTTS1 indicates objects
  $\Delta(I-K)<0.3$, WTTS2 indicates objects
  EW[Ca]$>1$\AA\  (see section~\ref{results1}),
  ALL indicates no $\Delta (I-K)$ restriction; (3)
  identifies which $\Delta R(q)$ model was used, r1 or r2 (see
  section~\ref{binary}); (4) identifies which $v_{\rm true}$
  distribution was assumed, v1 or v2 (see section~\ref{results2}); (5)
  the assumed value of $i_{\rm th}$ (see section 3.2); (6) the assumed value of the binary
  fraction $f$ (see section~\ref{binary}); (7) the assumed value of
  uncertainty in the observed (log) radius (see section~3.1); (8) the
  average value of $(\sin i)_{\rm obs}$; (9) the average value of
  $\alpha$; (10) the distance implied by equation~8 and its 1-sigma
  statistical uncertainty; (11) the Kolmogorov-Smirnov null-hypothesis
  probability (the probability that the observed distribution is not
  drawn from the model distribution is $1-P$(K-S)).}
\begin{tabular}{ccccccccccc}
\hline
Analysis No. & Sample (n) & $\Delta R(q)$ & $v_{\rm true}$ & $i_{\rm th}$  & bfrac & $\delta_{\log R}$ & $\langle (\sin
i)_{\rm obs} \rangle$ & $\langle \alpha \rangle$ & $D$ & $P$(K-S) \\
    &  & model & model  & degrees & $f$     & (dex)             & & &(pc) & \\                      
\hline
1 & WTTS1 (34) & r1 & v1 & 30 & 0.5 & 0.11 & 0.691 & 0.828 & $392\pm 23 $ & 0.66 \\
2 & WTTS1 (34) & r1 & v1 & 15 & 0.5 & 0.11 & 0.691 & 0.802 & $405\pm 24 $ & 0.75 \\
3 & WTTS1 (34) & r1 & v1 & 45 & 0.5 & 0.11 & 0.691 & 0.870 & $373\pm 22 $ & 0.61 \\
4 & WTTS1 (34) & r1 & v1 & 30 & 0.3 & 0.11 & 0.691 & 0.858 & $378\pm 22 $ & 0.67 \\
5 & WTTS1 (34) & r1 & v1 & 30 & 0.7 & 0.11 & 0.691 & 0.798 & $407\pm 24 $ & 0.63 \\
6 & WTTS1 (34) & r1 & v1 & 30 & 0.5 & 0.07 & 0.691 & 0.812 & $400\pm 18 $ & 0.64 \\
7 & WTTS1 (34) & r1 & v1 & 30 & 0.5 & 0.15 & 0.691 & 0.851 & $382\pm 28 $ & 0.41 \\
8 & WTTS1 (34) & r1 & v2 & 30 & 0.5 & 0.11 & 0.691 & 0.815 & $398\pm 24 $ & 0.64 \\
9 & WTTS1 (34) & r2 & v1 & 30 & 0.5 & 0.11 & 0.691 & 0.849 & $383\pm 22 $ & 0.68 \\  
10& WTTS2 (44) & r1 & v1 & 30 & 0.5 & 0.11 & 0.671 & 0.829 & $381\pm 20 $ & 0.66 \\  
11& ALL (74)   & r1 & v1 & 30 & 0.5 & 0.11 & 0.780 & 0.834 & $440\pm 19 $ & 0.48 \\ 
\hline
\end{tabular}
\label{resultstable}
\end{table*}

The results are collected in Table~\ref{resultstable}, and
Fig.~\ref{kscomp} demonstrates the sensitivity of the model
distribution to changes in the various parameters in the form of both
the differential and cumulative distributions of $\alpha (D/470)$
compared with those of the data.

Analysis number 1 from Table~\ref{results} is used as a baseline for the
discussion as this uses the parameters that could be considered to 
most likely 
represent the modelled ONC population. The average value of $(\sin
i)_{\rm obs}$ is significantly lower than would be expected for 
randomly inclined axes in agreement with the analysis of Rhode et
al. (2001) based on a smaller sample.
The simplest explanation is that the ONC is much closer than the
470\,pc assumed by Hillenbrand (1997). Indeed $D=470$\,pc is rejected
with 97 per cent confidence. I find a distance of 392\,pc
gives a much better representation of the data as may be judged from
Figs.~\ref{kscomp}a and~\ref{kscomp}b. The statistical
uncertainty is $\pm 23$\,pc which compares well with simulations and
uncertainty estimates presented by O'Dell et al. (1993) for similar
sized samples.

In addition to the statistical uncertainties I have tried other
models to test the sensitivity of the derived distance to variations in
model parameters that are only partially constrained. Fig.~\ref{kscomp}
shows examples of the model distributions at a fixed distance of
392\,pc in order to highlight these differences. The reader should note
that in all cases the variations in the intrinsic $\alpha$ distribution
can be compensated for by changes in the cluster distance to yield a
reasonable value for $P$(K-S) (these are the results quoted in
Table~\ref{resultstable}).  Therefore the observed $(\sin i)_{obs}$
distribution is
currently incapable of constraining some of the more interesting
parameters like $i_{\rm th}$, due to the small sample size and
experimental errors.

Figs.~\ref{kscomp}c and~\ref{kscomp}d show how changes of
$15^{\circ}<i_{\rm th}<45^{\circ}$ affect the modelled distribution at
a fixed distance of 392\,pc.  The changes are small and easily
compensated for by modest changes in the cluster distance of only $\pm
4.0$ per cent. This range seems a reasonable estimate of the
uncertainties engendered by the inclination bias -- if $i_{\rm
th}<15^{\circ}$ this would make little further difference to the
distance estimate and if $i_{\rm th}>45^{\circ}$ then there would be
difficulties in measuring rotation periods in a large fraction of WTTS.

Figs.~\ref{kscomp}e and~\ref{kscomp}f show the effects of changing the
binary frequency between $0.3<f<0.7$ at a fixed distance. Again the
small differences can be compensated for by small changes ($\pm 3.7$
per cent) in distance -- larger $f$ leads to larger distances.

Figs.~\ref{kscomp}g and~\ref{kscomp}h show that changing the assumed
value of $\delta_{\log R}$ in the plausible range $0.07<\delta_{\log R}<0.15$
significantly changes the shape of the distribution at a fixed
distance. This can be partially compensated for by changing the
distance by $\pm 2.3$ per cent. The broader distribution is a poorer
fit to the cumulative distribution function but cannot be rejected.

Figs.~\ref{kscomp}i and~\ref{kscomp}j show that even taking implausibly
large variations in the $v_{\rm true}$ and $\Delta R(q)$ distributions
(see sections~\ref{results2} and~\ref{binary} respectively) results in
very small changes in the model that can be compensated for by changes
in the cluster distance of only $\pm 1,9$ per cent.

For completeness, I show the effects of considering the sample of 74
objects (ALL) including both CTTS and WTTS (but all with spectral types of M2
or hotter) and also a sample of WTTS defined on the basis of EW[Ca]
(WTTS2 -- see section~\ref{results1}). As expected from Figs.~\ref{siniteff}
and~\ref{siniteff2} and the discussion in section~\ref{results1} the
distribution for the entire sample is skewed to higher $(\sin i)_{\rm
obs}$ values, resulting in a significantly larger distance estimate
with slightly better statistical precision ($440\pm 19$\,pc). The
sample of WTTS based on EW[Ca] is larger than sample WTTS1 but filters
out a few more of the objects with large $(\sin i)_{\rm obs}$
values. Hence the deduced distance based on sample WTTS2 is marginally
closer than that based on sample WTTS1.

The ``best'' final ONC distance estimate is 392\,pc. The total uncertainty
is 32\,pc, estimated from quadratic sum of the statistical uncertainty
and the distance uncertainties due to the parameter variations
discussed above. This of course assumes these uncertainties are
independent, ignores any systematic problem with the spectral
type-effective temperature relationship, and assumes that it is correct
to exclude CTTS from the analysis. The equivalent figure including the
CTTS would be $440\pm 34$\,pc, where the uncertainty in this case is
dominated by systematic effects.

\section{Discussion and Conclusions}

The observed $\sin i$ distribution of a group of WTTS (and CTTS) in the
ONC has been modelled.  The $\sin i$ values are linearly dependent on
the cluster distance and from this a distance to the ONC of $392\pm
32$\,pc is derived or $440\pm 34$\,pc if CTTS are included in the
sample.  The uncertainties include 1-sigma statistical uncertainties
and the contributions from uncertainties in a number of other partially
constrained model parameters such as the binary frequency, bias against
the inclusion of low inclination objects, the form of the true
equatorial velocity distribution and uncertanties in the stellar radius
estimates. With the present analysis and observational sample (34
WTTS), statistical and systematic uncertainties are of equal
importance. 

The observed $\sin i$ values depend linearly on the estimated radii,
which are difficult to determine for PMS stars. Hillenbrand (1997)
discusses these difficulties and concluded that extinctions,
luminosities and hence radii may be underestimated for accreting CTTS,
resulting in overestimates of $\sin i$ and hence overestimates of the
distance using the analysis presented in this paper. This effect is
readily apparent in the data, even when using imperfect discriminators
of accretion activity such as the $I-K$ excess and EW of the Ca\,{\sc
ii} 8542\AA\ line. A more careful analysis could use mid-IR
measurements to exclude (accretion) discs, but for now it seems wisest
to discard the suspected CTTS and adopt the closer of the two distance
estimates above.

A further issue is the validity of the spectral-type/effective
temperature relationship. As stated in section~\ref{results1}, the
relationship of Cohen \& Kuhi (1979) is uncontroversial in that it is
consistent with several later studies. However, these are also
fundamentally based on the atmospheres, angular diameters and
bolometric luminosities of cool main sequence stars. Young PMS stars
are likely to have different atmospheres and the fact that rotational
modulation has been measured in our sample stars means that they must
be covered by cool spotted regions. Assigning a single temperature to
such magnetically active stars may be difficult. These difficulties
could be worse in CTTS with their attendant discs, veiling and
accretion hotspots and this is a further reason for excluding them from
the analysis.  As yet, there is no compelling evidence for a change in
the spectral-type/temperature relationship for non-accreting PMS stars,
but as deduced distances will scale as $T_{\rm eff}^2$ this issue
should be borne in mind.

The deduced ONC distance is smaller than most previous results, although generally
agrees within the error bars.  If proper account of systematic
uncertainties is included, the distance derived here is also probably
more accurate than most previous determinations.  It is worth noting
however the excellent agreement with the $419\pm 21$\,pc determined by
Stassun et al (2004) from an eclipsing binary that they considered to
be a possible ONC member.

To recover the more usually used longer ONC distance of
450\,pc or 470\,pc (e.g. Hillenbrand 1997; Preibisch et al. 2005)
from the rotational data would require a model
with a very high binary fraction ($f\simeq 1$) combined with a
several per cent change in the relationship between spectral type and
temperature for warm K--M2 ONC stars.
A closer distance reduces deduced luminosities in the ONC by a
corresponding factor. Perhaps more importantly, ages deduced from the
positions of PMS stars in the HR diagram are increased by a factor of
$\simeq 1.7$, because $L \propto t^{-2/3}$ on fully convective PMS tracks.

The distance determined here is independent of stellar evolutionary
models -- an important source of systematic error in HR diagram-based
estimates. However, it does hinge on the assumption that the rotation
axes of PMS stars are randomly oriented in space. The observed $\sin i$
distribution is certainly consistent with this hypothesis, but is also
sufficiently broadened by experimental uncertainties that other
hypotheses cannot be rejected. For example, taking the parameters of
model 1 in Table~\ref{resultstable} and using a fixed inclination
$i=60^{\circ}$ results in a distance of 396\,pc and $P$(K-S)$=0.48$.

The theoretical and observational justification for random axial
orientation is mixed. In the quasi-static picture of star formation,
cloud collapse occurs preferentially parallel to magnetic field lines.
Some polarization studies suggest that disc (and
therefore presumably rotation) axes could be aligned with the large scale
magnetic fields in star forming regions (e.g. Vink et al. 2005). On the
other hand, imaging of CTTS discs in the Taurus-Auriga region reveals
a random orientation with respect to the large scale magnetic field, 
in accord with a more dynamical star formation picture
(M\'enard \& Duch\^ene 2004).  

As a final remark, previous applications of the rotationally based
technique to the Pleiades cluster and Taurus-Auriga association have
yielded distances which turn out to be in good agreement with later
precise parallax-based measurements (see O'Dell et al. 1994; Preibisch
\& Smith 1997). This tends to argue that the underlying assumptions of
the method (random axial orientations and a dwarf-like temperature
scale for non-accreting PMS stars) are reasonable.  A subsequent
precise trigonometric parallax for the ONC (e.g. using the Hubble Space
Telescope) could be used to test the robustness of these model
assumptions in more detail.

\section{Acknowledgements}
I thank William Herbst for some very useful comments on the initial
version of this paper.

\nocite{vink05}
\nocite{menard04}
\nocite{breger81}
\nocite{hillenbrand97}
\nocite{hillenbrand98}
\nocite{penston75}
\nocite{hendry93}
\nocite{odell94}
\nocite{herbst05}
\nocite{herbst02}
\nocite{herbst00}
\nocite{stassun99}
\nocite{schussler96}
\nocite{johnskrull97}
\nocite{neuhauser98}
\nocite{preibisch97}
\nocite{joncour94}
\nocite{warren78}
\nocite{slesnick04}
\nocite{lada00}
\nocite{muench02}
\nocite{huff06}
\nocite{shuping06}
\nocite{odell98}
\nocite{walker69}
\nocite{genzel81}
\nocite{bertout99}
\nocite{anthonytwarog82}
\nocite{rhode01}
\nocite{sicilia05}
\nocite{cohen79}
\nocite{joneswalker88}
\nocite{stassun06}
\nocite{stassun04}
\nocite{prosser94}
\nocite{padgett97}
\nocite{petr98}
\nocite{simon99}
\nocite{kohler06}
\nocite{fischer92}
\nocite{duquennoy91}
\nocite{kenyon95}
\nocite{leggett96}
\nocite{bessell88}
\nocite{siess00}
\nocite{baraffe02}
\nocite{rebull06}

\bibliographystyle{mn2e}  
\bibliography{iau_journals,master}

\begin{thebibliography}{}

\bibitem[\protect\citeauthoryear{{Anthony-Twarog}}{{Anthony-Twarog}}{1982}]{an%
thonytwarog82}
{Anthony-Twarog} B.~J.,  1982, AJ, 87, 1213

\bibitem[\protect\citeauthoryear{Baraffe, Chabrier, Allard \&
  Hauschildt}{Baraffe et~al.}{2002}]{baraffe02}
Baraffe I.,  Chabrier G.,  Allard F.,    Hauschildt P.~H.,  2002, A\&A, 382,
  563

\bibitem[\protect\citeauthoryear{{Bertout}, {Robichon} \& {Arenou}}{{Bertout}
  et~al.}{1999}]{bertout99}
{Bertout} C.,  {Robichon} N.,    {Arenou} F.,  1999, A\&A, 352, 574

\bibitem[\protect\citeauthoryear{{Bessell} \& {Brett}}{{Bessell} \&
  {Brett}}{1988}]{bessell88}
{Bessell} M.~S.,  {Brett} J.~M.,  1988, PASP, 100, 1134

\bibitem[\protect\citeauthoryear{{Breger}, {Gehrz} \& {Hackwell}}{{Breger}
  et~al.}{1981}]{breger81}
{Breger} M.,  {Gehrz} R.~D.,    {Hackwell} J.~A.,  1981, ApJ, 248, 963

\bibitem[\protect\citeauthoryear{{Cohen} \& {Kuhi}}{{Cohen} \&
  {Kuhi}}{1979}]{cohen79}
{Cohen} M.,  {Kuhi} L.~V.,  1979, ApJS, 41, 743

\bibitem[\protect\citeauthoryear{Duquennoy \& Mayor}{Duquennoy \&
  Mayor}{1991}]{duquennoy91}
Duquennoy A.,  Mayor M.,  1991, A\&A, 248, 485

\bibitem[\protect\citeauthoryear{Fischer \& Marcy}{Fischer \&
  Marcy}{1992}]{fischer92}
Fischer D.~A.,  Marcy G.~W.,  1992, ApJ, 396, 178

\bibitem[\protect\citeauthoryear{{Genzel}, {Reid}, {Moran} \&
  {Downes}}{{Genzel} et~al.}{1981}]{genzel81}
{Genzel} R.,  {Reid} M.~J.,  {Moran} J.~M.,    {Downes} D.,  1981, ApJ, 244,
  884

\bibitem[\protect\citeauthoryear{{Hendry}, {O'dell} \&
  {Collier-Cameron}}{{Hendry} et~al.}{1993}]{hendry93}
{Hendry} M.~A.,  {O'dell} M.~A.,    {Collier-Cameron} A.,  1993, \mnras, 265,
  983

\bibitem[\protect\citeauthoryear{{Herbst}, {Bailer-Jones}, {Mundt},
  {Meisenheimer} \& {Wackermann}}{{Herbst} et~al.}{2002}]{herbst02}
{Herbst} W.,  {Bailer-Jones} C.~A.~L.,  {Mundt} R.,  {Meisenheimer} K.,
  {Wackermann} R.,  2002, \aap, 396, 513

\bibitem[\protect\citeauthoryear{{Herbst} \& {Mundt}}{{Herbst} \&
  {Mundt}}{2005}]{herbst05}
{Herbst} W.,  {Mundt} R.,  2005, \apj, 633, 967

\bibitem[\protect\citeauthoryear{{Herbst}, {Rhode}, {Hillenbrand} \&
  {Curran}}{{Herbst} et~al.}{2000}]{herbst00}
{Herbst} W.,  {Rhode} K.~L.,  {Hillenbrand} L.~A.,    {Curran} G.,  2000, \aj,
  119, 261

\bibitem[\protect\citeauthoryear{Hillenbrand}{Hillenbrand}{1997}]{hillenbrand9%
7}
Hillenbrand L.~A.,  1997, AJ, 113, 1733

\bibitem[\protect\citeauthoryear{{Hillenbrand}, {Strom}, {Calvet}, {Merrill},
  {Gatley}, {Makidon}, {Meyer} \& {Skrutskie}}{{Hillenbrand}
  et~al.}{1998}]{hillenbrand98}
{Hillenbrand} L.~A.,  {Strom} S.~E.,  {Calvet} N.,  {Merrill} K.~M.,  {Gatley}
  I.,  {Makidon} R.~B.,  {Meyer} M.~R.,    {Skrutskie} M.~F.,  1998, \aj, 116,
  1816

\bibitem[\protect\citeauthoryear{{Huff} \& {Stahler}}{{Huff} \&
  {Stahler}}{2006}]{huff06}
{Huff} E.~M.,  {Stahler} S.~W.,  2006, ApJ, 644, 355

\bibitem[\protect\citeauthoryear{{Johns-Krull} \& {Hatzes}}{{Johns-Krull} \&
  {Hatzes}}{1997}]{johnskrull97}
{Johns-Krull} C.~M.,  {Hatzes} A.~P.,  1997, ApJ, 487, 896

\bibitem[\protect\citeauthoryear{{Joncour}, {Bertout} \& {Bouvier}}{{Joncour}
  et~al.}{1994}]{joncour94}
{Joncour} I.,  {Bertout} C.,    {Bouvier} J.,  1994, A\&A, 291, L19

\bibitem[\protect\citeauthoryear{{Jones} \& {Walker}}{{Jones} \&
  {Walker}}{1988}]{joneswalker88}
{Jones} B.~F.,  {Walker} M.~F.,  1988, AJ, 95, 1755

\bibitem[\protect\citeauthoryear{Kenyon \& Hartmann}{Kenyon \&
  Hartmann}{1995}]{kenyon95}
Kenyon S.~J.,  Hartmann L.~W.,  1995, ApJS, 101, 117

\bibitem[\protect\citeauthoryear{{K{\"o}hler}, {Petr-Gotzens}, {McCaughrean},
  {Bouvier}, {Duch{\^e}ne}, {Quirrenbach} \& {Zinnecker}}{{K{\"o}hler}
  et~al.}{2006}]{kohler06}
{K{\"o}hler} R.,  {Petr-Gotzens} M.~G.,  {McCaughrean} M.~J.,  {Bouvier} J.,
  {Duch{\^e}ne} G.,  {Quirrenbach} A.,    {Zinnecker} H.,  2006, A\&A, 458, 461

\bibitem[\protect\citeauthoryear{{Lada}, {Muench}, {Haisch} Jr., {Lada},
  {Alves}, {Tollestrup} \& {Willner}}{{Lada} et~al.}{2000}]{lada00}
{Lada} C.~J.,  {Muench} A.~A.,  {Haisch} Jr. K.~E.,  {Lada} E.~A.,  {Alves}
  J.~F.,  {Tollestrup} E.~V.,    {Willner} S.~P.,  2000, AJ, 120, 3162

\bibitem[\protect\citeauthoryear{Leggett, Allard, Berriman, Dahn \&
  Hauschildt}{Leggett et~al.}{1996}]{leggett96}
Leggett S.~K.,  Allard F.,  Berriman G.,  Dahn C.~C.,    Hauschildt P.~H.,
  1996, ApJS, 104, 117

\bibitem[\protect\citeauthoryear{{M{\'e}nard} \& {Duch{\^e}ne}}{{M{\'e}nard} \&
  {Duch{\^e}ne}}{2004}]{menard04}
{M{\'e}nard} F.,  {Duch{\^e}ne} G.,  2004, \aap, 425, 973

\bibitem[\protect\citeauthoryear{{Muench}, {Lada}, {Lada} \& {Alves}}{{Muench}
  et~al.}{2002}]{muench02}
{Muench} A.~A.,  {Lada} E.~A.,  {Lada} C.~J.,    {Alves} J.,  2002, ApJ, 573,
  366

\bibitem[\protect\citeauthoryear{{Neuh\"{a}user}, {Wolk}, {Torres},
  {Preibisch}, {Stout-Batalha}, {Hatzes}, {Frink}, {Wichmann}, {Covino},
  {Alcala}, {Brandner}, {Walter}, {Sterzik} \& {Koehler}}{{Neuh\"{a}user}
  et~al.}{1998}]{neuhauser98}
{Neuh\"{a}user} R.,  {Wolk} S.~J.,  {Torres} G.,  {Preibisch} T.,
  {Stout-Batalha} N.~M.,  {Hatzes} A.~P.,  {Frink} S.,  {Wichmann} R.,
  {Covino} E.,  {Alcala} J.~M.,  {Brandner} W.,  {Walter} F.~M.,  {Sterzik}
  M.~F.,    {Koehler} R.,  1998, \aap, 334, 873

\bibitem[\protect\citeauthoryear{{O'Dell}}{{O'Dell}}{1998}]{odell98}
{O'Dell} C.~R.,  1998, AJ, 115, 263

\bibitem[\protect\citeauthoryear{{O'Dell}, {Hendry} \& {Collier
  Cameron}}{{O'Dell} et~al.}{1994}]{odell94}
{O'Dell} M.~A.,  {Hendry} M.~A.,    {Collier Cameron} A.,  1994, \mnras, 268,
  181

\bibitem[\protect\citeauthoryear{{Padgett}, {Strom} \& {Ghez}}{{Padgett}
  et~al.}{1997}]{padgett97}
{Padgett} D.~L.,  {Strom} S.~E.,    {Ghez} A.,  1997, ApJ, 477, 705

\bibitem[\protect\citeauthoryear{{Penston}, {Hunter} \& {O'Neill}}{{Penston}
  et~al.}{1975}]{penston75}
{Penston} M.~V.,  {Hunter} J.~K.,    {O'Neill} A.,  1975, \mnras, 171, 219

\bibitem[\protect\citeauthoryear{{Petr}, {Coude Du Foresto}, {Beckwith},
  {Richichi} \& {McCaughrean}}{{Petr} et~al.}{1998}]{petr98}
{Petr} M.~G.,  {Coude Du Foresto} V.,  {Beckwith} S.~V.~W.,  {Richichi} A.,
  {McCaughrean} M.~J.,  1998, \apj, 500, 825

\bibitem[\protect\citeauthoryear{{Preibisch} \& {Smith}}{{Preibisch} \&
  {Smith}}{1997}]{preibisch97}
{Preibisch} T.,  {Smith} M.~D.,  1997, \aap, 322, 825

\bibitem[\protect\citeauthoryear{{Prosser}, {Stauffer}, {Hartmann},
  {Soderblom}, {Jones}, {Werner} \& {McCaughrean}}{{Prosser}
  et~al.}{1994}]{prosser94}
{Prosser} C.~F.,  {Stauffer} J.~R.,  {Hartmann} L.,  {Soderblom} D.~R.,
  {Jones} B.~F.,  {Werner} M.~W.,    {McCaughrean} M.~J.,  1994, ApJ, 421, 517

\bibitem[\protect\citeauthoryear{{Rebull}, {Stauffer}, {Megeath}, {Hora} \&
  {Hartmann}}{{Rebull} et~al.}{2006}]{rebull06}
{Rebull} L.~M.,  {Stauffer} J.~R.,  {Megeath} S.~T.,  {Hora} J.~L.,
  {Hartmann} L.,  2006, ApJ, 646, 297

\bibitem[\protect\citeauthoryear{{Rhode}, {Herbst} \& {Mathieu}}{{Rhode}
  et~al.}{2001}]{rhode01}
{Rhode} K.~L.,  {Herbst} W.,    {Mathieu} R.~D.,  2001, AJ, 122, 3258

\bibitem[\protect\citeauthoryear{{Sch\"ussler}, {Caligari}, {Ferriz-Mas},
  {Solanki} \& {Stix}}{{Sch\"ussler} et~al.}{1996}]{schussler96}
{Sch\"ussler} M.,  {Caligari} P.,  {Ferriz-Mas} A.,  {Solanki} S.~K.,    {Stix}
  M.,  1996, \aap, 314, 503

\bibitem[\protect\citeauthoryear{{Shuping}, {Kassis}, {Morris}, {Smith} \&
  {Bally}}{{Shuping} et~al.}{2006}]{shuping06}
{Shuping} R.~Y.,  {Kassis} M.,  {Morris} M.,  {Smith} N.,    {Bally} J.,  2006,
  ApJ, 644, L71

\bibitem[\protect\citeauthoryear{{Sicilia-Aguilar}, {Hartmann},
  {Hern{\'a}ndez}, {Brice{\~n}o} \& {Calvet}}{{Sicilia-Aguilar}
  et~al.}{2005}]{sicilia05}
{Sicilia-Aguilar} A.,  {Hartmann} L.~W.,  {Hern{\'a}ndez} J.,  {Brice{\~n}o}
  C.,    {Calvet} N.,  2005, AJ, 130, 188

\bibitem[\protect\citeauthoryear{Siess, Dufour \& Forestini}{Siess
  et~al.}{2000}]{siess00}
Siess L.,  Dufour E.,    Forestini M.,  2000, A\&A, 358, 593

\bibitem[\protect\citeauthoryear{{Simon}, {Close} \& {Beck}}{{Simon}
  et~al.}{1999}]{simon99}
{Simon} M.,  {Close} L.~M.,    {Beck} T.~L.,  1999, \aj, 117, 1375

\bibitem[\protect\citeauthoryear{{Slesnick}, {Hillenbrand} \&
  {Carpenter}}{{Slesnick} et~al.}{2004}]{slesnick04}
{Slesnick} C.~L.,  {Hillenbrand} L.~A.,    {Carpenter} J.~M.,  2004, ApJ, 610,
  1045

\bibitem[\protect\citeauthoryear{{Stassun} \& {Mathieu}}{{Stassun} \&
  {Mathieu}}{2006}]{stassun06}
{Stassun} K.,  {Mathieu} R.,  2006, in American Astronomical Society Meeting
  Abstracts {A Survey For Pre-main-sequence Spectroscopic Binary Stars In The
  Orion Nebula Cluster}.
p. 08.03

\bibitem[\protect\citeauthoryear{Stassun, Mathieu, Mazeh \& Vrba}{Stassun
  et~al.}{1999}]{stassun99}
Stassun K.~G.,  Mathieu R.~D.,  Mazeh T.,    Vrba F.~J.,  1999, AJ, 117, 2941

\bibitem[\protect\citeauthoryear{{Stassun}, {Mathieu}, {Vaz}, {Stroud} \&
  {Vrba}}{{Stassun} et~al.}{2004}]{stassun04}
{Stassun} K.~G.,  {Mathieu} R.~D.,  {Vaz} L.~P.~R.,  {Stroud} N.,    {Vrba}
  F.~J.,  2004, ApJS, 151, 357

\bibitem[\protect\citeauthoryear{{Vink}, {Drew}, {Harries}, {Oudmaijer} \&
  {Unruh}}{{Vink} et~al.}{2005}]{vink05}
{Vink} J.~S.,  {Drew} J.~E.,  {Harries} T.~J.,  {Oudmaijer} R.~D.,    {Unruh}
  Y.,  2005, \mnras, 359, 1049

\bibitem[\protect\citeauthoryear{{Walker}}{{Walker}}{1969}]{walker69}
{Walker} M.~F.,  1969, ApJ, 155, 447

\bibitem[\protect\citeauthoryear{{Warren} Jr. \& {Hesser}}{{Warren} \&
  {Hesser}}{1978}]{warren78}
{Warren} Jr. W.~H.,  {Hesser} J.~E.,  1978, ApJS, 36, 497

\end{thebibliography}


\bsp 

\label{lastpage}

\end{document}